\documentclass[usenatbib]{mn2e}
\usepackage{multirow}
\usepackage{rotating}
\usepackage{colortbl}
\usepackage{color}
\usepackage[fleqn]{amsmath}
\usepackage{amssymb}
\usepackage{amsfonts}
\usepackage{verbatim}
\usepackage{scalefnt}
\usepackage[percent]{overpic}

\usepackage[bookmarks,bookmarksnumbered,colorlinks=true,citecolor=blue,linkcolor=cyan]{hyperref}
\usepackage{cleveref}

\newcommand{\apjl}{ApJ}
\newcommand{\apjs}{ApJS}
\newcommand{\ao}{Appl.~Opt.}
\newcommand{\aap}{A\&A}

\newcommand{\beq}{\begin{equation}}
\newcommand{\eeq}{\end{equation}}

\usepackage{soul,xcolor}

\definecolor{grey}{rgb}{0.5,0.6,0.7}
\def \simlt { \lower .75ex \hbox{$\sim$} \llap{\raise .27ex \hbox{$<$}} }
\definecolor{purple}{rgb}{0.65,0.15,0.9}
\definecolor{darkorange}{rgb}{0.8,0.3,0}
\definecolor{olive}{rgb}{0.4,0.6,0.25}
\definecolor{darkgreen}{rgb}{0,0.7,0}
\definecolor{darkred}{rgb}{0.5,0,0}

\title[Properties of simulated galaxies and BHs in cosmic voids]{Properties of simulated galaxies and supermassive black holes in cosmic voids}
\author[Habouzit et al.]{M\'{e}lanie Habouzit$^{1,2,3}$\thanks{E-mail: habouzit@mpia-hd.mpg.de},
Alice Pisani$^{4}$, Andy Goulding$^{4}$,
Yohan Dubois$^{5}$, 
\newauthor
Rachel S. Somerville$^3$,
and Jenny E. Greene$^4$  \\
 $^1$ Max-Planck-Institut für Astronomie, Konigstuhl 17, D-69117, Heidelberg, Germany\\
 $^2$ Zentrum für Astronomie der Universität Heidelberg, Albert-Ueberle-Str. 2, D-69120 Heidelberg, Germany\\
 $^3$ Center for Computational Astrophysics, Flatiron Institute, New York, NY 10010, USA\\
 $^4$ Department of Astrophysical Sciences, Princeton University, 4 Ivy Lane, Princeton, NJ, 08544, USA\\
 $^5$ Institut d’Astrophysique de Paris, Sorbonne Universit\'es, CNRS, UMR 7095, 98 bis bd Arago, 75014 Paris, France\\
 }

\begin{document}
\maketitle

\begin{abstract}

Cosmic voids, the under-dense regions of the cosmic web, are widely used to constrain cosmology. Voids contain few, isolated galaxies, presumably expected to be less evolved and preserving memory of the pristine Universe.
We use the cosmological hydrodynamical simulation Horizon-AGN coupled to the void finder {\sc \texttt{VIDE}} to investigate properties of galaxies in voids at $z=0$. 
We find that, closer to void centers, low-mass galaxies are more common than their massive counterparts. 
At fixed dark matter halo mass, they have smaller stellar masses than in denser regions.
The star formation rate of void galaxies diminishes when approaching void centers, but their sSFR slightly increases, suggesting that void galaxies form stars more efficiently with respect to their stellar mass.
We find that this can not only be attributed to the prevalence of low-mass galaxies.
The inner region of voids also predominantly host low-mass BHs. 
However, the BH mass to galaxy mass ratios resemble those of the whole simulation at $z=0$. 
Our results suggest that even if the growth channels in cosmic voids are different than in denser environments, voids grow their galaxies and BHs in a similar way. While a large fraction of the BHs have low Eddington ratios, we find that $\sim 20\%$ could be observed as AGN with $\log_{10} L_{\rm 2-10 \, keV}=41.5-42.5 \, \rm erg/s$.
These results pave the way to future work with larger next-generation hydro simulations, aiming to confirm our findings and prepare the application on data from upcoming large surveys such as PFS, Euclid and WFIRST.

\end{abstract}


\section{Introduction}
\label{sec:intro}
The evolution of the primordial density perturbations led to the formation of the large-scale structure pattern that we see today---dubbed the cosmic web---and composed of dark matter halos, large galaxy clusters, dense filaments, and emptier regions. These under-dense regions of the cosmic web---known as cosmic voids---have sizes from 10 to 100 of $h^{-1}\, \rm Mpc$. They have been widely used in recent years to extract cosmological information \citep[see, e.g.][and references therein]{2015A&C.....9....1S,2015PhRvD..92h3531P,2016PhRvL.117i1302H,2017JCAP...07..014H,2017A&A...607A..54H,2019arXiv190305161P}. Voids have a simple dynamic \citep[they are dominated by coherent single-stream motions, see e.g.][]{2011JCAP...05..015S,2016PhRvL.117i1302H}. They are thus expected to be less evolved and preserve memory of the initial Universe. In this landscape, one of the topics of high interest is the study of galaxy properties and evolution in cosmic voids.

While the understanding of galaxy evolution already involves very complex physical processes, considering galaxy mergers and external processes (e.g., stripping or harassment, generally called {\it nurture}) further raises the level of complexity. The evolution of galaxies in the central regions of voids is more likely to be driven by in-situ processes only ({\it nature}), where none to very few galaxy mergers happen. Therefore cosmic voids would offer a unique laboratory of less evolved environments where one can study galaxy evolution in the absence of mergers and associated physical processes, and consequently understand the role of environment on galaxy evolution. 
Observationally, void galaxies have shown different properties than galaxies in more average or denser regions of the Universe \citep{2004ApJ...617...50R,2005ApJ...624..571R,2006MNRAS.372.1710P,2008MNRAS.384.1189V,2012MNRAS.426.3041H,2012AJ....144...16K,2014MNRAS.439.3564C}, but are still debated today \citep{2006MNRAS.372.1710P,2012AJ....144...16K} and the conclusions driven so far may be related to the high relative abundance of low-mass galaxies in voids, as explained below. 
Statistically, it has been found that galaxies embedded in cosmic voids are bluer \citep{1999AJ....118.2561G,2004ApJ...617...50R}, and tend to have later morphological type \citep{1999AJ....118.2561G,2004ApJ...617...50R}.
From observations, it seems that these void galaxies also have stellar disks with smaller radii compared to a control sample of late-type galaxies \citep{2011ASSP...27...17V}.
Using galaxy samples from SDSS, \citet{2005ApJ...624..571R} compare the equivalent width (H$\alpha$, [OII], [NII], H$\beta$, [OIII]) of galaxies located in voids or walls. Larger equivalent widths are found in void galaxies, indicating that they are forming stars at a high rate, while having smaller masses, and being fainter. It results in a higher specific star formation rate \citep[sSFR, see also][and references therein]{2011ASSP...27...17V}. The same conclusions were found in \citep{2004ApJ...617...50R}, using photometric properties.
Recently, the analysis of galaxy evolution in SDSS DR7, including more than 6000 void galaxies, showed that galaxies located in cosmic voids evolve slowly with respect to the general galaxy population, with almost all void galaxies still being on the galaxy star-forming sequence (also called {\it main sequence}) at $z\sim0.1-0.01$ \citep{2014MNRAS.445.4045R}. 
The gas metallicity of void galaxies has been subject to debate.
Void galaxies also seems to have lower gas metallicity \citep{2011AstBu..66..255P}, but these results are based on samples with only a very few galaxies, and sometimes they are part of the same parent and larger void, which may cause some bias \citep[see discussion in][for which no significant difference is found for the metallicity]{2014arXiv1410.6597K}. 
Whether these void galaxies have indeed different properties than galaxies located in denser environments, or if these differences are due or partly due to the low mass bias is still subject to intense debate. 
This paper aims at providing a theoretical framework to start answering these questions. Hydrodynamical cosmological simulations are a powerful resource when it comes to studying statistically and self-consistently the evolution of thousands of galaxies, especially as a function of their environment. The primary goal of this paper is to study the mass and star-forming properties of galaxies located in cosmic voids.  \\

If, as suggested by these works using observational samples, galaxies in the central region of voids are indeed less evolved or ``late'' compared to galaxies in walls or denser regions, these galaxies may be ideal candidates to look for clues of pristine supermassive BH properties.
BH formation in the early Universe is far from being understood, and can unfortunately not be assessed with our current observational facilities. BHs are thought to play a predominant role in galaxy formation: they are present in very different types of galaxies from dwarf galaxies to massive elliptical \citep{2012NatCo...3E1304G,2013ARA&A..51..511K,2013ApJ...764..184M,2013ApJ...775..116R,2015ApJ...799...98M}, and are believed to drive powerful feedback into their host galaxies and surroundings shaping the massive end of the galaxy luminosity function \citep[][and references therein]{Silk2013}.

The tightest observational constraint on BH formation mechanisms comes from the observation of high-redshift quasars, which are powered by the most luminous BHs, at $z=6-7$ \citep{Mortlock2011}. These $\sim 10^{8} \, \rm M_{\odot}$ objects tell us that BH seeds must have formed in the early Universe \citep[see][for a review, and references therein]{2017PASA...34...31V} with initial mass in the range $100-10^{5}\, \rm M_{\odot}$. 
Theories of BH formation include the remnants of the first generation of stars \citep[{\it PopIII remnant model}, $M_{\rm BH}\sim 100 \, \rm M_{\odot}$][]{MadauRees2001,VHM}, the collapse of very massive stars formed in primordial gas \citep[{\it Direct collapse model}, $M_{\rm BH}\sim 10^{4-6}\, \rm M_{\odot}$][]{LoebRasio1994,2003ApJ...596...34B,2004MNRAS.354..292K,2006ApJ...652..902S,BVR2006,LN2006,2008MNRAS.391.1961D,Wise2008,Regan09}, and the collapse of massive stars formed by runaway collisions in compact stellar clusters \citep[{\it compact stellar cluster model}, $M_{\rm BH}\sim 10^{3} \, \rm M_{\odot}$][]{2008ApJ...686..801O,Devecchi2009,Regan09,Katz2015}. 
These models determine the initial mass of BH seeds, and their abundance (i.e., the probability for galaxies to host a BH).

Clues to distinguish between these mechanisms are not to be found in evolved and massive galaxies, where several galaxy-galaxy and BH-BH mergers, as well as increased dynamical interaction, have certainly erased BH initial properties (i.e., BH mass and BH occupation fraction). 
All theoretical BH formation models yield distinctive galaxy occupation fractions, i.e., different probabilities that a galaxy of a given mass hosts a BH \citep{2008MNRAS.383.1079V,2010MNRAS.408.1139V,2012NatCo...3E1304G}. 
Since low-mass galaxies experience fewer mergers, their BH occupation is one diagnostic for BH formation models \citep{2012NatCo...3E1304G}. Void galaxies, if mainly driven by in-situ processes, could provide a way of shedding new light on the ``pristine'' BH occupation fraction. We investigate this aspect in a forthcoming paper (Habouzit et al. in prep) and focus here our attention on the properties of BHs in void galaxies. The mass of BHs in low-mass galaxies is the second diagnostics to distinguish between the different formation mechanisms.
The possible elevated number of low-mass galaxies expected in voids could bring new constraints on the initial mass of BHs, as well as what drives their growth from birth until today.

Finally, the fact that observed void galaxies are mostly on the main sequence \citep{2014MNRAS.445.4045R} may consequently indicate that their central BHs (if they exist) may be still active as well. This has been shown in early studies of SDSS DR2, where AGN were found to be common in void galaxies \citep{2008ApJ...673..715C}. It could favor the observations of such BHs in low-mass void galaxies, or at least provide regions of the Universe to look for low-mass systems. The presence of AGN in voids is still debated today, as in the Void Galaxy Survey (VGS) sample in which only 1 AGN has been identified in a early-type galaxy over 59 galaxies  \citep{2016MNRAS.458..394B,2017MNRAS.464..666B}.
Studying BH growth in cosmic voids can also constrain the contribution of the different growth channels, i.e., BH gas accretion and BH-BH mergers. BH mergers are likely to be sub-dominant in void environments. BHs located in the inner region of cosmic voids are also less likely to be impacted by large cold accretion flows from large-scale filaments. Voids represent an excellent opportunity to study the role of {\it nature} versus {\it nurture} for BH growth.

The present paper aims at studying the properties of galaxies and their BHs in cosmic voids using the state-of-the-art cosmological hydrodynamical simulation Horizon-AGN. To date, theoretical studies of the properties of galaxies and BHs in cosmological hydrodynamical simulations have remained limited to individuals void regions, e.g. using zoom-in simulation method \citep{2015ApJ...799..178K}. While this allows to analyze a few regions in details carefully, it does not provide a cosmological view of the diversity of void environments neither a statistical understanding of the galaxy and BH properties in voids.

The simulation Horizon-AGN has been able to reproduce a large range of galaxy properties \citep{2014MNRAS.444.1453D,2016MNRAS.463.3948D,2017MNRAS.467.4739K}, and BH properties \citep{2016MNRAS.460.2979V}, and is, therefore, suitable for the present investigation.
Several reasons motivate the analyses presented in the following.
First, we are in a crucial period: while galaxy properties in cosmic voids have been studied in broad observational surveys as SDSS, for which the voids have large sizes of few tens cMpc, studying their properties in large-scale cosmological hydrodynamical simulations has been neglected. In state-of-the-art simulations such as the one we use in this paper, of side length 100 cMpc, the sizes of cosmic voids are smaller than in some of the SDSS analyses (we discuss this aspect in more details in the paper).
The next-generation hydro cosmological simulations have increased volume, as the IllustrisTNG with 300 cMpc side length \citep{2018MNRAS.473.4077P}. They will allow for the identification of voids with sizes and statistics similar to those of the SDSS surveys.
The potential benefit of cosmic voids to study and constrain BH formation has been neglected as well. Observationally the presence of AGN in voids is unclear; in this paper we provide predictions on the expected population of BHs in cosmic voids.
Our present analysis, although based on smaller voids than in observations, paves the way for future investigations. In this context and because the paper lies at the edges of several domains such as the physics of voids, galaxies and BHs, we describe in details the technique that we use to identify voids in the simulation, and the physical motivation for each analysis. We also discuss in further detail the perspectives of such analyses and ongoing other investigations at the end of the paper.
Finally, we also point out that the next generation of galaxy surveys, such as the Prime Focus Spectrograph (PFS, $15 \rm deg^{2}$ of the sky, and $1400\, \rm deg^{2}$ for the wide field), the Euclid mission ($15000\, \rm deg^{2}$ for the full spectroscopic survey, deep field of $40 \, \rm deg^{2}$), and the Wide Field Infrared Survey Telescope (WFIRST, $2200\, \rm deg^{2}$ for the full spectroscopic survey, $\sim 10\, \rm deg^{2}$ for the deep field), will be crucial to enable new progress for our understanding of galaxy properties in cosmic voids \citep{2019arXiv190305161P}. The new mission SPHEREx promises also to be a powerful tool to study the large-scale structures by mapping the entire sky ($\sim 40000 \, \rm deg^{2}$) soon, with a focus at relatively low redshift ($z<0.6$).
We describe the simulation Horizon-AGN and its relevant physical models in Section 2, and the void finder code in Section 3. We present the galaxy evolution in voids in Section 4, and BH evolution and properties in Section 5.

\section{Cosmological simulations Horizon-AGN}
\subsection{{\sc Ramses} code and initial conditions}
\label{subsec:ramses}

In this paper, we analyze the large scale state-of-the-art cosmological hydrodynamical simulation Horizon-AGN \citep{2014MNRAS.444.1453D}, which was run with the adaptive mesh refinement (AMR) code {\sc Ramses } \citep{2002A&A...385..337T}. 
Particles are projected on the grid with a Cloud-In-Cell interpolation, and the Poisson equation is solved with an adaptive Particle-Mesh solver.
The Euler equations are solved with a second-order unsplit Godunov scheme using an approximate HLLC Riemann solver, with a Min-Mod total variation diminishing scheme to interpolate the cell-centered values to their edge locations.
We refine the initial mesh with seven levels of refinement to achieve a spatial resolution of $\Delta x=1\, \rm kpc$.
Cells are refined/unrefined based on a quasi-Lagrangian criterion if there are more/less than 8 DM particles in a cell, or if the total baryonic mass is higher/smaller than 8 times the DM mass resolution. 
New level of refinement are only allowed when the expansion factor double (i.e., $a_{\rm exp}=0.1,0.2,0.4$) to keep the refinement homogeneous in physical units with cosmic time. \\
The simulation uses a standard $\Lambda$ cold dark matter cosmology, with parameters compatible with those of the {\it Wilkinson Microwave Anisotropy Probe} (WMAP7) parameters\footnote{These cosmological parameters are compatible within 10 per cent relative variation with {\it Planck} parameters \citep{planck2013}.}\citep{Spergel2007}: total matter density $\Omega_{m}=0.272$, dark matter energy density $\Omega_{\Lambda}=0.728$, amplitude of the matter power spectrum $\sigma_{8}=0.81$, spectral index $\rm{n_{s}}=0.967$, baryon density $\Omega_{b}=0.045$ and Hubble constant $\rm{H_{0}}= 70.4\, \rm{km\,  s^{-1} \, Mpc^{-1}}$. 

\subsection{Physics of the simulation}
Here, we summarize the main sub-grid models of the simulations. All details can be found in \citet{2014MNRAS.444.1453D}.\\

\noindent {\bf Radiative cooling and photoheating by UV background}\\
Radiative cooling is modeled with the cooling curves of \cite{1993ApJS...88..253S}; the gas cools down to $10^{4}\, \rm K$ through H, He, with a contribution from metals.
To mimic reionization, photoheating from a uniform ultraviolet radiation background is added, taking place after redshift z=10, following \citet{1996ApJ...461...20H}.
Gas metallicity is modeled as a passive variable,  
and an initial metallicity background of ${\rm Z}=10^{-3}\, \rm Z_{\odot}$ is used for the two simulations. The metallicity is modified by the injection of gas ejecta from SN explosions and stellar winds. The simulations account for the release of several chemical elements synthesized in stars and released by stellar winds and SNe: O, Fe, C, N, Mg, and Si, but these elements do not contribute separately to the cooling curve. \\

\noindent {\bf Star formation and SN feedback}\\
Star formation occurs in dense ($\rho>\rho_{0}$, with $\rho$ the density of the gas, $\rho_{0}=0.1\, \rm{H\, cm}^{-3}$ the gas hydrogen number density threshold, see equation~\ref{eq:SF}) and cold ($T<T_{0}$, see equation~\ref{eq:TSF}) cells. 
The model follows the Kennicutt-Schmidt law:
\begin{equation}
\frac{d\rho_{\star}}{dt}=\epsilon_{\star}\frac{\rho}{t_{\rm{ff}}},
\label{eq:SF}
\end{equation}
with $\dot{\rho_{\star}}$ the star formation rate density, $\epsilon_{\star}=0.02$ the star formation efficiency (constant with redshift), $\rho$ the density of the gas and $t_{\rm{ff}}$ the free-fall time of the gas. 
The probability of forming N stars with a mass resolution of  $m_{\rm{res,\star}}=\rho_{0}\Delta x^{3}\sim 2\times 10^{6}\, M_{\rm \odot}$ is computed with a Poisson random process.\\
The gas follows an adiabatic equation-of-state (EoS) for monoatomic gas with adiabatic index  $\gamma=5/3$. At high gas densities ($\rho>\rho_0$) we add a polytropic EoS to increase the gas pressure and to limit excessive gas fragmentation by mimicking heating of the interstellar medium from stars~\citep{springel&hernquist03}:
\begin{equation}
T=T_{0}\left(\frac{\rho}{\rho_{0}}\right)^{\kappa-1},
\label{eq:TSF}
\end{equation}
with $T$ the gas temperature, $T_{0}$ the temperature threshold, $\rho_{0}$ the density threshold, and $\kappa=4/3$ the polytropic index of the gas. \\
Stellar SN feedback is modeled as kinetic winds and assumes a Salpeter initial mass function with a low-mass cut-off of $0.1 \, \rm M_{\odot}$ and a high-mass cut-off of $100 \, \rm M_{\odot}$.\\ 

\noindent {\bf Black hole formation and AGN feedback}\\
BHs are modeled as collisionless sink particles; they accrete surrounding gas and merge together. The Horizon-AGN seeding model forms BHs of fixed mass $M_{\rm seed}=10^{5}\,M_{\rm \odot}$ (less if there is not enough gas in the cell) in dense regions ($\rho>\rho_{0}$)
BH accretion follows the Bondi-Hoyle-Lyttleton model:
\begin{eqnarray}
\dot{M}_{\rm BH}=\min\left(\frac{4\pi \alpha G^{2} M_{\rm BH}^{2}\bar{\rho} }{(\bar{c}^{2}+\bar{v}^{2})^{3/2}},\dot{M}_{\rm Edd}\right),
\end{eqnarray}
with 
\begin{eqnarray}
\dot{M}_{\rm Edd}=\frac{4\pi G M_{\rm BH} m_{\rm p}}{\epsilon_{\rm r} \sigma_{T} c},
\end{eqnarray}
where $\alpha$ is a boost factor \citep{Booth2009} ($\alpha=(\rho/\rho_{0})^{2}$ for $\rho>\rho_{0}$, and $\alpha=1$ otherwise), G is the gravitational constant, $M_{\rm BH}$ the mass of the BH, $\bar{\rho}$ the average density of the medium, $\bar{c}$ the average of the sound speed, $\bar{v}$ the average velocity of the gas relative to the BH, $m_{\rm p}$ the proton mass, $\epsilon_{\rm r}=0.1$ the radiative efficiency, and $\sigma_{\rm T}$ the Thomson cross-section.

BHs can impact their host galaxies through AGN feedback. The suite of these two simulations Horizon-AGN and Horizon-noAGN is a powerful tool to analyze the impact of AGN feedback. AGN feedback is a combination of two modes \citep[see][for a complete description of the implementation]{2012MNRAS.420.2662D}: radio mode for low BH accretion rates when $\dot{M}_{\rm BH}/\dot{M}_{\rm Edd}<0.01$, and quasar mode for high BH accretion rates with $\dot{M}_{\rm BH}/\dot{M}_{\rm Edd}>0.01$. 
The radio mode deposits AGN feedback energy into a bipolar cylindrical outflow with a jet velocity of $10^{4}\, \rm km s^{-1}$, with an energy of $\dot{E}_{\rm AGN}=\epsilon_{\rm f} \epsilon_{\rm r} \dot{M}_{\rm BH}c^{2}$, with an efficiency $\epsilon_{\rm f}=1$. 
Instead, the quasar mode is modeled as an isotropic injection of thermal energy into the gas within a sphere of radius $\Delta x$ with an energy of $\dot{E}_{\rm AGN}=\epsilon_{\rm f} \epsilon_{\rm r} \dot{M}_{\rm BH}c^{2}$, and an efficiency $\epsilon_{\rm f}=0.15$. The efficiency parameters are set to reproduced the scaling relations between BH mass and galaxy properties \citep{2012MNRAS.420.2662D}. The quasar mode takes place after an episode of high gas accretion rates onto the BHs, thus mostly in high-redshift galaxies, which are gas rich and provide a gas reservoir to feed the BHs.

\subsection{Dark matter halo, galaxy and BH catalogs}
Dark matter halos and sub-halos are identified using AdaptaHOP halo finder~\citep{2004MNRAS.352..376A} ({\sc HaloMaker}). AdaptaHOP uses an SPH-like kernel to compute densities at the location of each particle and partitions the ensemble of particles into sub-halos based on saddle points in the density field. A total of 20 neighbors are used to compute the local density of each particle, and we use a density threshold of $\Delta_{\rm c}=178$ times the average total matter density. 
The force softening (minimum size below which substructures are considered irrelevant) is $\sim 2\,\rm{kpc}$. The dark matter resolution is $M_{\rm DM}=8\times 10^{7}\, \rm M_{\odot}$, and only dark matter halos with more than 50 dark matter particles are considered here ($M_{\rm DM}\geqslant 5\times 10^{9}\, \rm M_{\odot}$). 
We use the same scheme (but based on the stellar distribution) to identify galaxies. We only consider galaxies with stellar masses of $M_{\rm gal}\gtrsim 2\times 10^{8}\, \rm M_{\odot}$, i.e. including at least 50 stellar particles.
We assign BHs to their host galaxies based on mass and position: the most massive BH within a galaxy virial radius is assigned to that galaxy.


\begin{figure*}
\centering
\includegraphics[scale=0.85]{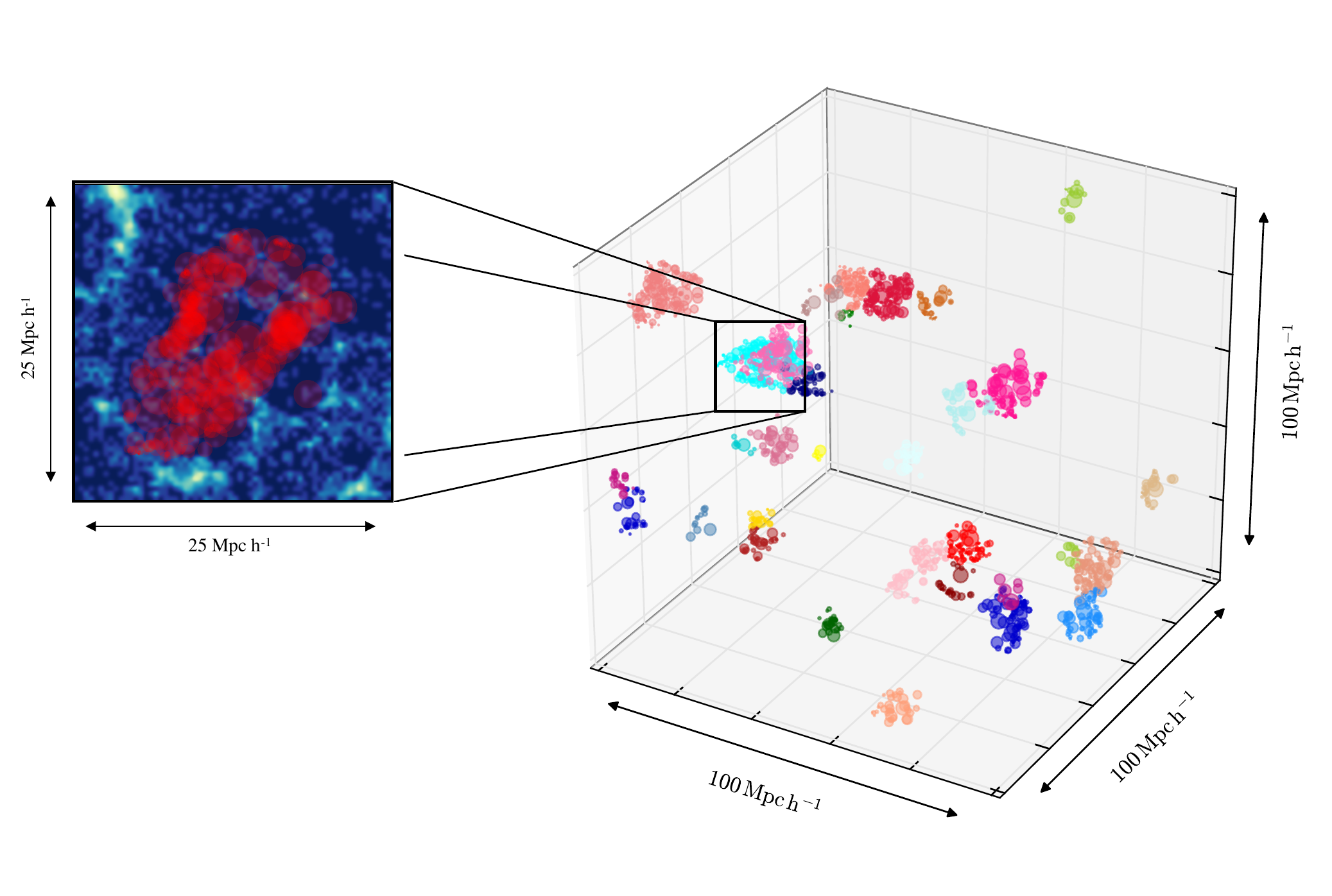}
\caption{{\it Right panel}: Three-dimensional representation of the simulated volume of Horizon-AGN (axes in $\rm cMpc/h$). Colored circles indicate the position of the galaxies considered as included in a given cosmic void (we only show galaxies within $0.8 R_{\rm void}$), galaxies of the same colors are within the same void. The size of the circles is proportional to the volume of the Voronoi tessellation cell in which a given galaxy is embedded. The top left void provides an example of void structure, with the most isolated galaxies at the center of the void (large circles), and non-isolated galaxies (small circles for small Voronoi cell volume) at the edges. These latter galaxies are likely to be considered as wall or shell galaxies in our analysis.
{\it Left panel}: Zoom into a void region where galaxies belonging to the void are highlighted with red circles. These circles are proportional to the galaxy Voronoi cell volume. The background represents the density of the tracers, here the simulated galaxies, light blue regions having a higher density of galaxies. }
\label{fig:simu_volume}
\end{figure*}

\section{Finding void galaxies in Horizon-AGN}
\label{sec:section_voids}
\subsection{Identifying cosmic voids and void galaxies}
Cosmic voids are becoming an increasingly important part of large-scale structure analysis. For this reason, many algorithms exist to construct void catalogs from both data and simulations. Some confusion may arise because very different void finders exist, and despite the common goal of selecting low-density areas of the cosmic web, they often follow distinct procedures and hence deal in the end with different objects. In this paper, we focus on 3D voids found in the tracer distribution, which can either be galaxies, dark matter particles, or dark matter halos.

We can divide 3D void finders into three main categories \citep[][and references therein for specific examples]{2008MNRAS.387..933C,2010MNRAS.403.1392L,2018MNRAS.476.3195C}. 
First, void finders can be based on the tracer distribution to define the density field. This family of void finders uses tracers to select under-dense (not empty) regions---and hence can provide a density profile for such regions. The second family also relies on tracers but will select empty spheres in their distribution. 
Finally, the last kind of void finder algorithms exploits the dynamical information and defines voids as outflow dominated regions. Cosmological analysis in recent works mainly relies on the two first categories \citep[see e.g.][]{2016PhRvL.117i1302H,2017JCAP...07..014H,2017A&A...607A..54H,2018arXiv180606860P}.

For the purpose of our work, since we need to study galaxy and BH properties in regions of low density, it would be essential to identify voids that are under-dense but not empty. We thus use a void finder that falls into the first category and selects regions that are under-dense in the cosmic web; hence these zones include tracers in isolated regions. We now present the details of the algorithm we use in this work, and we briefly discuss how we mitigate possible impacts of void definition when performing the selection of void galaxies, by exploiting different outputs of the algorithm.

To analyze the properties of galaxies and BH in voids, we use the \texttt{VIDE} toolkit\footnote{\url{https://bitbucket.org/cosmicvoids/vide\_public/wiki/Home}} \citep{2015A&C.....9....1S}, a public code based on \texttt{ZOBOV} \citep{2008MNRAS.386.2101N}. 
The void finding procedure relies on a  Voronoi Tessellation of the tracer distribution: for each galaxy, a cell is constructed such that any point in the cell is closer to that galaxy. Using galaxies as tracers here is convenient since it mimics what is done in observations. 
Interestingly, the method allows obtaining an estimate of the local density by considering the reciprocal of the cell volume as an estimate for the local value of the density field. We will see that this feature is particularly useful for the purposes of our paper. The density minima in the field (corresponding to the largest Voronoi cells) are then used to create catchment basins by joining cells surrounding each minimum. 
The tessellation and minimum definition step that provides basins are followed by the application of the Watershed Transform  \citep{2007MNRAS.380..551P} to construct under-dense regions: it builds a full void hierarchy by merging basins under the condition that the ridge between them is lower than 20\% of the mean density of the sample. 
The void finding procedure defines the center of the void as the volume-weighted barycenter, and the void radius (an effective radius) is then given by 
\begin{equation}
R_{\rm{void}}\equiv\bigg( \frac{3}{4\pi}V_{\rm gal}\bigg)^{1/3}
\end{equation}
where the volume $V_{\rm gal}$ is the total volume of the Voronoi cells composing the void. 
Since \texttt{ZOBOV} (and hence \texttt{VIDE}) constructs voids that contain tracers (such as galaxies)---and does not simply find voids by considering empty regions---it is thus possible to analyze the properties of galaxies in the obtained under-dense regions.

It is important to stress that \texttt{VIDE} includes within voids the high-density regions surrounding those voids, commonly known as walls. 
It means that, if we use voids defined by \texttt{VIDE}, we can use the radius information to analyze galaxy properties when approaching the under-dense central regions of voids. Of course, relying on the void radius means that we will be averaging out the shape information\footnote{This effect could be mitigated when using larger samples of voids.} (as void are not spherical on a one to one basis); but it also allows us to see an evolution of features with void radius clearly.
Typically we will be able to see how galaxy properties evolve from the under-dense central region of voids to the higher-density region at the void's edge. The effect can be more easily studied thanks to the possibility of stacking voids of different sizes by rescaling them in radius. Hence even though that we average out the void shape information, using void radius provides us with the advantage of analyzing galaxies within regions of similar under-dense properties. We can indeed observe how galaxy and SMBH properties evolve as a function of distance from the center of the void regardless of the size of the voids. 
Additionally, the center definition relies on the volume weighted barycenter procedure---which means that it bears information about the walls---mitigating the fact that defining a center in regions of low tracer density can be noise dominated and reduce the precision of the center determination. 

This is the first paper investigating the behavior of galaxies within voids with such a large hydrodynamical simulation.\\ 

With this in mind, two possible approaches will be used throughout this paper to study BHs and galaxies in under-dense regions.

\noindent {\bf Void galaxies defined with void-centric distance}\\
First, it is possible to consider voids found through the full procedure---i.e., voids belonging to the hierarchy that can be defined in the cosmic web---and look for the properties of galaxies in the central and most under-dense regions of such voids. It can be done using the radius $R_{\rm void}$ information. It has the disadvantage of relying on an average measure (voids have very different shapes, the radius of voids being an average quantity, it only partially captures void properties), but allows to stack information based on the void size and enhance the signal over noise by consistently scaling and overlapping voids of different sizes. This method of stacking voids is often used in observational studies to understand the global properties of galaxies in voids as a function of their locations within these voids.\\

\noindent {\bf Void galaxies defined with Voronoi cell volume}\\
The second approach is to directly focus on the Voronoi cells by selecting the cells with higher volume (and thus corresponding to the most isolated galaxies). In this framework, the Voronoi tessellation procedure is particularly effective as it provides a quick way to select the most isolated galaxies in the simulation. Interestingly, this second approach can also provide direct access to denser parts of the simulation, thus serving as a valuable comparison with the under-dense regions. It would correspond to look at properties of galaxies near the wall of the void (at $r/R_{\rm void} \simeq 1$ for spherical voids), without any prior assumption on where the cell is. Also, using the Voronoi tessellation information allows to extract information directly using the distance of the cells from the void center instead of the void radius, thus not assuming any particular shape of the void nor spherically averaging such distance (hence, conversely to the above radius-based case).\\

As both approaches present advantages, we decide to follow both of them in this paper, aiming to understand which one is more efficient and reliable to prepare the analysis of real data (such as an SDSS-based void catalog, for which the same void finder has already been used \citep[e.g.,][]{2016PhRvL.117i1302H,2017JCAP...07..014H}). Therefore in the following, we describe the properties of galaxies and their BHs as a function of void-centric distances and Voronoi cell volumes.

\begin{figure}
\centering
\includegraphics[scale=0.51]{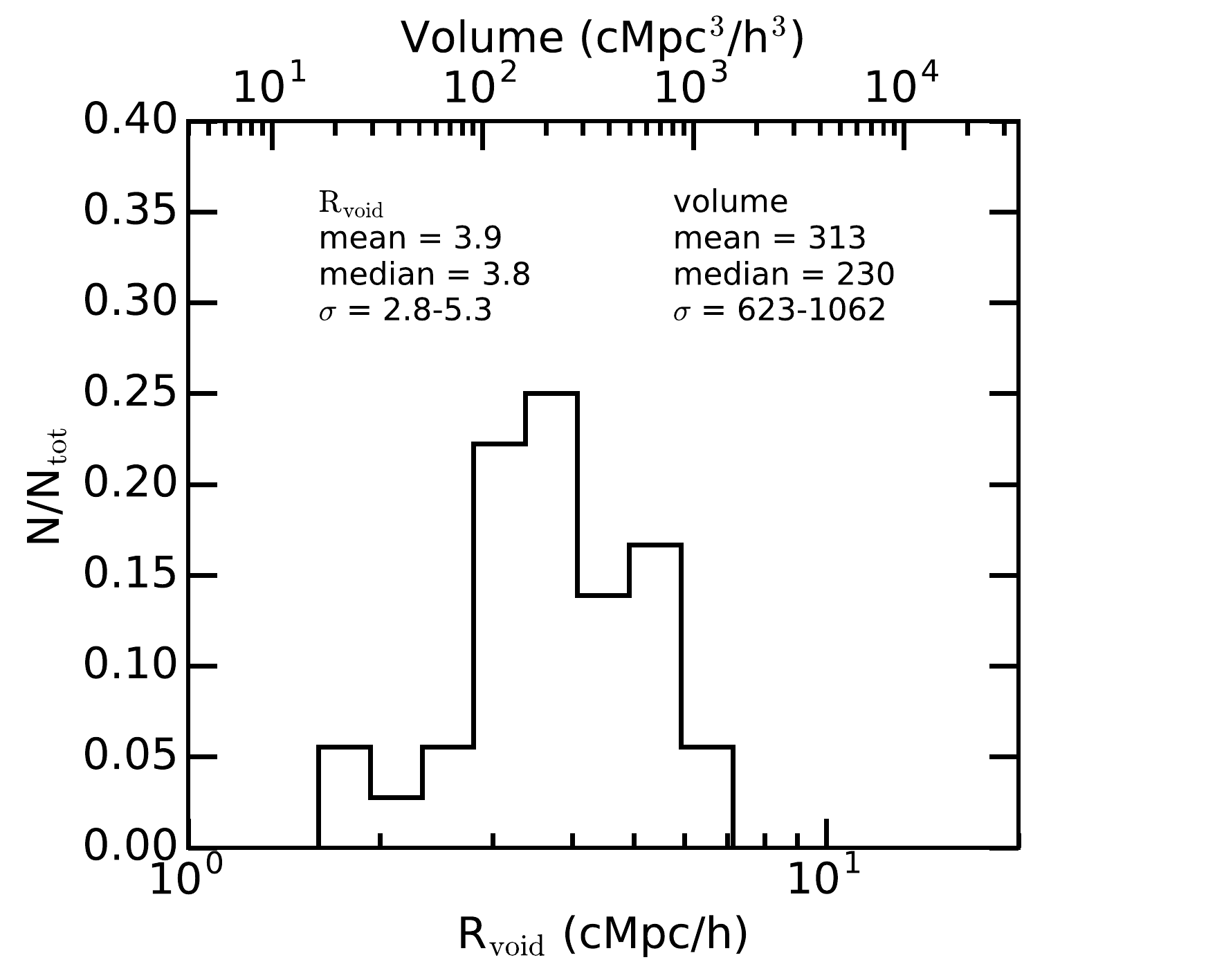}
\caption{Distribution of the radius (bottom x-axis) and volume (top x-axis) of the voids identified in Horizon-AGN. The voids have a median radius of 3.8 $\rm cMpc/h$ and volume of $230 \, \rm (cMpc/h)^{3}$. We also indicate the mean of the distributions, as well as their standard deviation $\pm \sigma$.}
\label{fig:stat_void_distri}
\end{figure}

\subsection{Void galaxy catalog from Horizon-AGN}

We describe in this section the void catalog that we have computed for Horizon-AGN at redshift $z=0$.
In Fig.~\ref{fig:simu_volume}, we show the 3D simulation box of Horizon-AGN, where we highlight the galaxies located in cosmic voids as colored circles, whose radii are equal to twice the volume of their Voronoi tessellation cells derived with the void finder code. The 36 cosmic voids identified are shown in different colors.
We show one of these cosmic voids in the left side panel, where again galaxies belonging to the void are shown as red circles. 
As explained below, we have made conservative choices to select our sample of voids.

In Fig.~\ref{fig:stat_void_distri}, we show the size distribution of the voids identified in Horizon-AGN. 
We find that the voids in Horizon-AGN have a median radius of $\tilde{R}_{\rm void}= 3.8\, \rm cMpc/h$ and a median volume of $\tilde{V}_{\rm void}=230\, \rm (cMpc/h)^{3}$. These voids are smaller than the voids identified in observations. For comparison, the voids identified in the SDSS DR7 survey considered in \citet{2014MNRAS.445.4045R} have radius $\geqslant 10 \, \rm cMpc/h$.
The resolution of the simulation is higher than current observational surveys. Therefore, we have access here to what we could call subvoids. Some of these small under-dense regions might be embedded in larger regions that can only be identified in larger simulated volume or observational surveys. The simulation allows us to describe the low-level structure of the cosmic web with great precision.
Our work paves the way to access the information that we could get from upcoming deep-field surveys.

\begin{figure}
\centering
\includegraphics[scale=0.51]{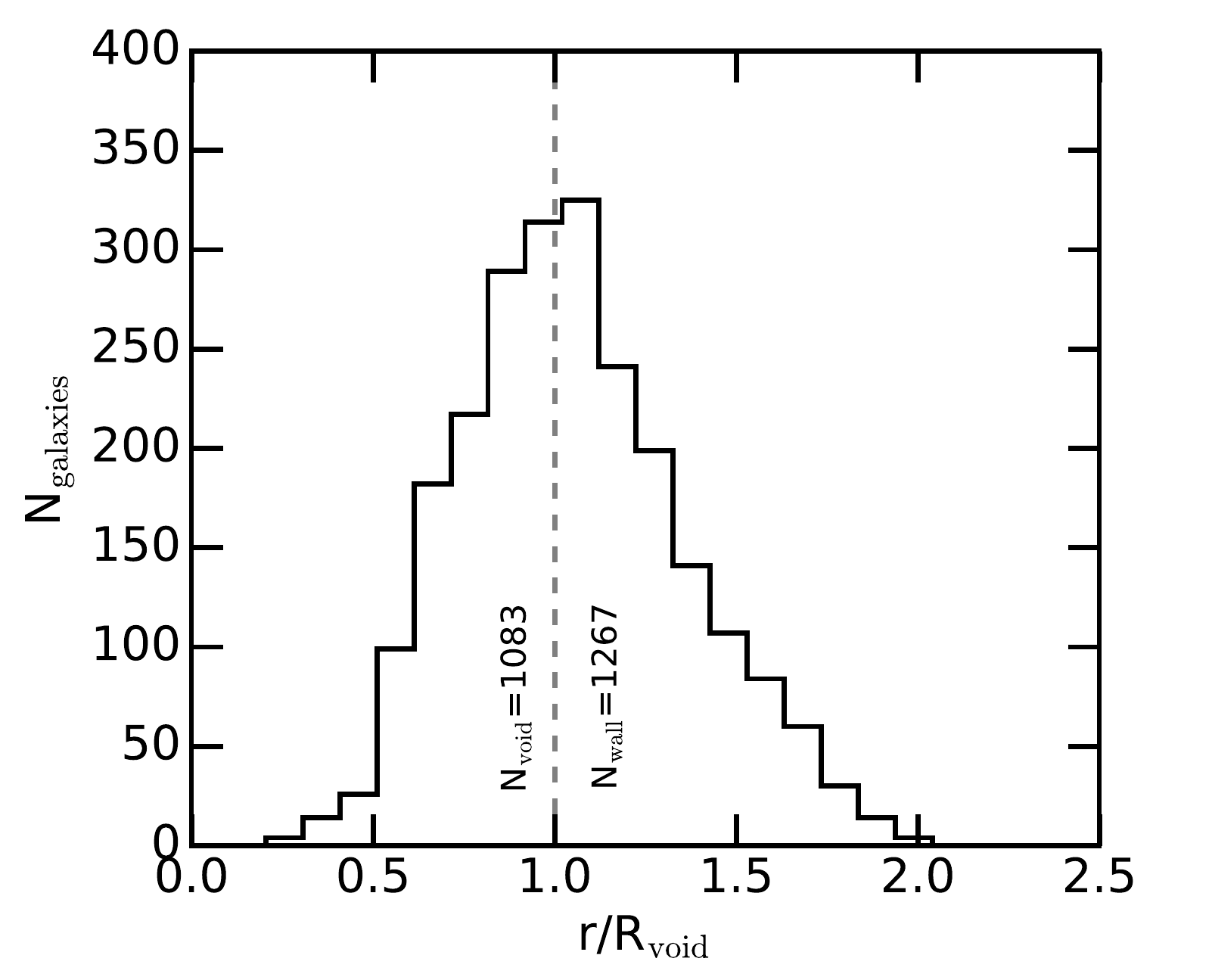}
\caption{Number of galaxies in voids as a function of their void-centric distance, normalized to the void radius $R_{\rm void}$. The dash vertical grey line indicates the radius of the voids. There are 1083 void galaxies with $r/R_{\rm void}<1$ in our catalog, which represents about $1\%$ of the total number of galaxies in the simulation at that time. The distribution vanishes after $r/R_{\rm void}\sim1$ because we only include galaxies that are linked to voids in our catalog and not all the galaxies of the simulation.}
\label{fig:stat_void}
\end{figure}

In Fig.~\ref{fig:stat_void}, we show the number of galaxies as a function of normalized radius $r/R_{\rm void}$, i.e. void-centric distance of a galaxy $r$ normalized by the radius $R_{\rm void}$ of the void hosting this galaxy. We identified 1083 void galaxies defined by $r/R_{\rm void}\leqslant 1$ in our catalog, which represents less than $1\%$ of the total number of galaxies in the simulation volume at that time. The catalog also includes galaxies with $r/R_{\rm void}>1$, which are part of the denser regions around cosmic voids but are still linked to a given void by the void finder code\footnote{If voids are non-spherical shape, some of the $r/R_{\rm void}>1$ can actually be located inside the voids, and vice-versa.}. 
In the following we divide galaxies in three categories:
\begin{itemize}
\item void galaxies: defined as $r/R_{\rm void}\leqslant 1$, number of galaxies $N_{\rm void}=1083$, $94\%$ of the void galaxies are central galaxies\footnote{Galaxies are defined as central if most massive galaxies in their dark matter halos, and satellites otherwise.};  when specified, we also sometimes use a more conservative choice and refer to void galaxies as $r/R_{\rm void}\leqslant0.8$ in the following;
\item shell galaxies: $0.8 \leqslant r/R_{\rm void}\leqslant 1$, $N_{\rm shell}=584$, $93\%$ are central galaxies;
\item inner void galaxies: $r/R_{\rm void}\leqslant 0.65$, $N_{\rm inner}=195$, $96\%$ of the inner void galaxies are central galaxies.
\end{itemize}

We note that while the values chosen to separate categories are chosen for convenience, results are not impacted by the exact thresholds chosen, and are qualitatively robust. We empirically define shell galaxies with the limit $r/R_{\rm void}=0.8$, as most of the void galaxy properties (that we study in the following) statistically reach the average behavior of all the simulated galaxy sample at this $r/R_{\rm void}$ value. We hence note that, throughout the whole paper, when we refer to galaxies in the walls of voids (that is in the shell), we expect to find a similar average behaviour as for galaxies of the rest of the simulation.

\begin{figure}
\centering
\includegraphics[scale=0.51]{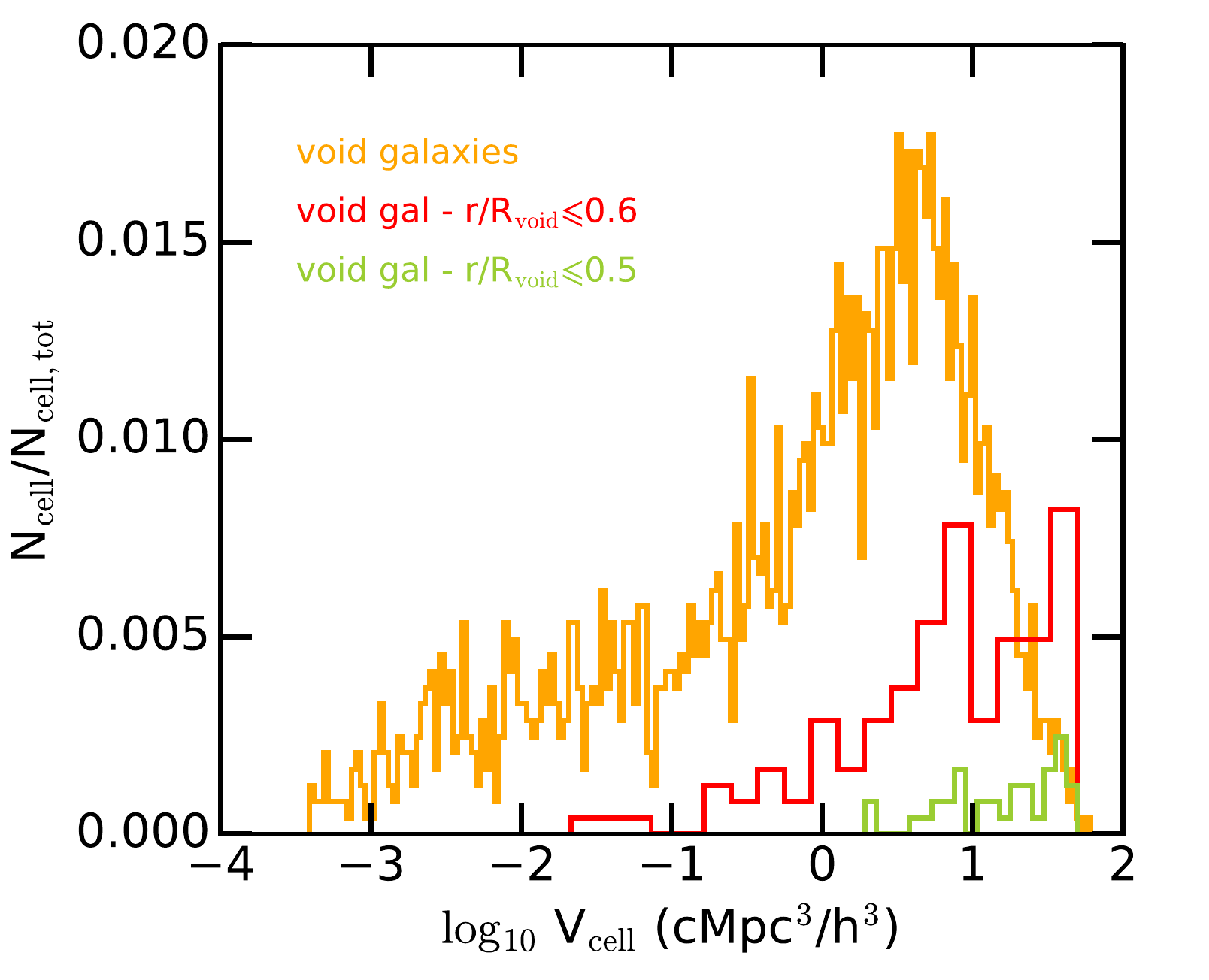}
\caption{Normalized distribution of Voronoi cell volume of galaxies in voids (orange histogram), and of void galaxies within $50\%$ and $60\%$ of the void radius. Void galaxies close to the center of void preferentially have larger Voronoi cell volume. This shows that galaxies in the inner region of voids, as defined with void-centric distances, are preferentially isolated galaxies with large Voronoi cell volume. We note some exceptions of $r/R_{\rm void}\leqslant 0.6$ galaxies that are not very isolated in the cosmic web, with a lower volume of their Voronoi cells.} 
\label{fig:stat_void_cellvolume}
\end{figure}

\subsection{Are the most isolated galaxies actually located in the center of cosmic voids?}
As explained above, in this paper, we will present galaxy and BH properties as a function of both void-centric radius, and volume of Voronoi cells around galaxies. To understand the connection between these two quantities, we show in Fig.~\ref{fig:stat_void_cellvolume} the distribution of Voronoi cell volume around void galaxies for several void-centric radius bins. We see that galaxies embedded in the inner region of voids with void-centric radii or $r/R_{\rm void}\leqslant 0.6$ (red distribution) correspond to larger Voronoi cell volume, and even more so for galaxies with $r/R_{\rm void}\leqslant 0.5$ (green distribution). The smaller the void-centric distance, the larger the cell volume. This is very instructive: it shows that galaxies in the central region of voids are also, on average, the most isolated galaxies. In theory, selecting galaxies as a function of their void-centric radius, or according to their large Voronoi cell volume should provide us with similar information. In practice, as we work here with a hydrodynamical cosmological simulation of $100\, \rm Mpc/h$ side length that does not contain many voids, using void-centric distances and stacking voids to study the properties of galaxies can lead to some noise in the signal. As the voids do not always have a spherical shape, some galaxies could be considered as living in the central region of void while in reality being at the edges. In this case, using the Voronoi cell volume as an indicator of isolation is essential. We will describe this effect on the different analyses in the following sections.


\section{Properties of galaxies in voids}
In this section, our primary objectives are to understand what are the properties of galaxies that populate the cosmic voids identified in Horizon-AGN. More specifically, we want to understand the mass properties of these galaxies (section 4.1), how they evolve within their dark matter halos (section 4.2), whether their evolution is driven by in-situ processes (section 4.3), and their ability to form stars (section 4.4). We also analyze how these properties evolve with void-centric distances.

\subsection{Galaxy mass properties in voids}

\begin{figure}
\centering
\includegraphics[scale=0.35]{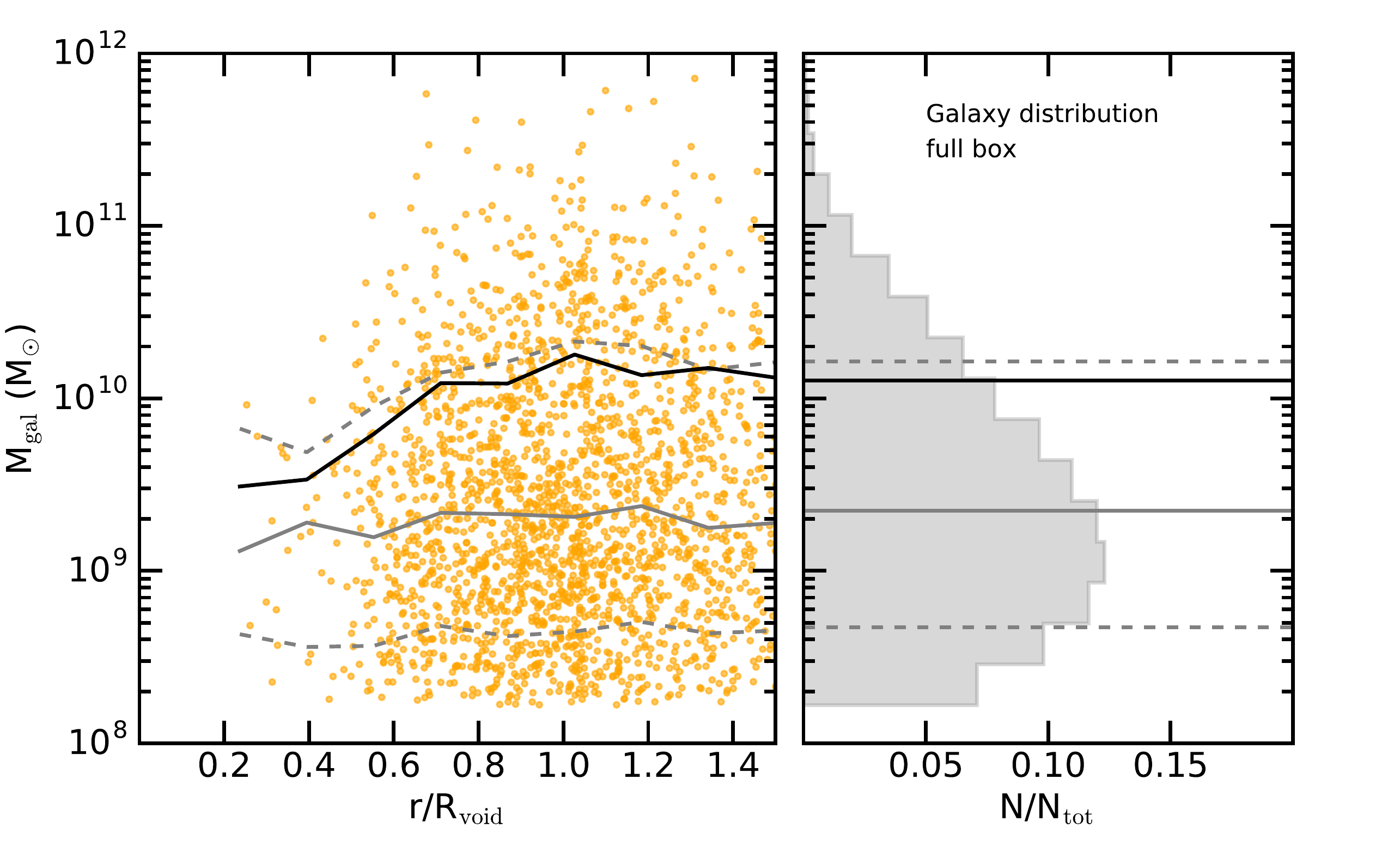}
\caption{Galaxy mass as a function of normalized void-centric radius $r/R_{\rm void}$. Grey distribution in the right panel shows the galaxy mass distribution for the full simulation box. On both panels, black lines indicate the mean of the distributions, the solid grey lines the median of the distributions, and the dashed grey lines the standard deviation $\pm \sigma$. Galaxies located in the inner regions of voids have lower total stellar mass $M_{\rm gal}$. 
}
\label{fig:distri_gal_voidcentric}
\end{figure}

In Fig.~\ref{fig:distri_gal_voidcentric}, we show the total stellar mass of void galaxies as a function of their void-centric distance in yellow dots. The solid black line indicates the mean of galaxy mass, the solid grey line the median of the distribution, and the dashed grey lines the standard deviation $\pm \sigma$ of the distribution. For comparison, we also provide the mass distribution of galaxies for the entire simulation on the right-side panel, with the same meanings for the different colored lines. 
The relative abundance of low-mass galaxies is higher in the center of voids.
Indeed, there are less and less massive galaxies when approaching the center of voids. The mean mass of galaxies in the simulation is reached at $r/R_{\rm void}\sim 0.8-1$, in good agreement with observational findings in SDSS galaxy void catalogs \citep{2014MNRAS.445.4045R}.

\begin{figure}
\centering
\includegraphics[width=\columnwidth]{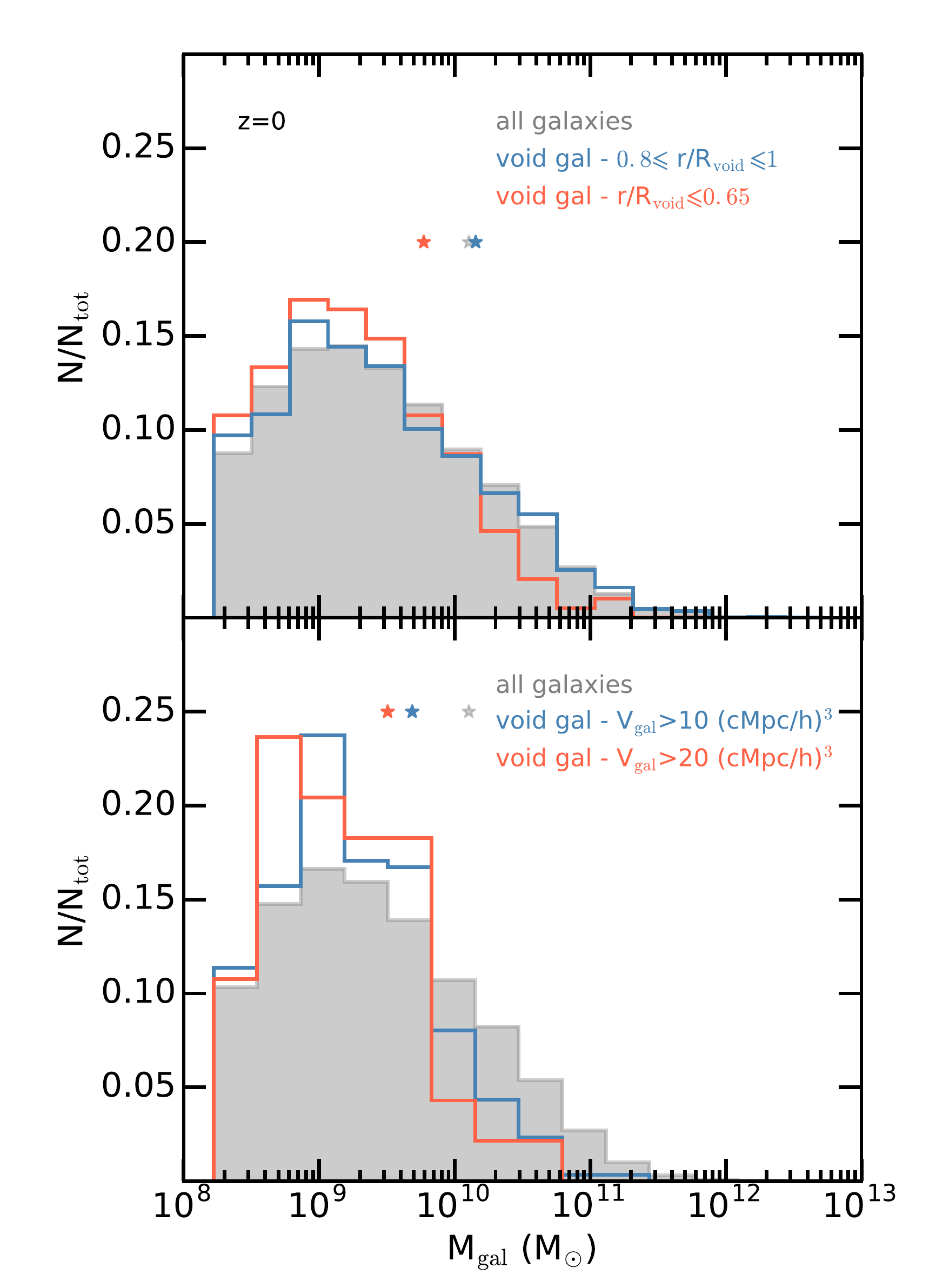}
\caption{Galaxy mass distribution for all galaxies, normalized to the number of galaxies, in grey shaded distribution in both panels. In the top panel, we also show the distribution of galaxy mass for galaxies located at the edges of voids ($0.80 \leqslant r/R_{\rm void} \leqslant 1$) in blue, and the distribution of inner galaxies (with $r/R_{\rm void}\leqslant 0.65$) in red. $N_{\rm tot} $ values for all samples are given in Section 3.2. In the bottom panel, we show the mass distributions for galaxies with a volume of their Voronoi cell higher than $\rm V_{cell}>10\, (cMpc/h)^{3}$ in blue, and the even more isolated galaxies with $\rm V_{cell}>20\, (cMpc/h)^{3}$ in red. The inner void galaxy and the isolated galaxy distributions are significantly distinct from the distribution of all simulated galaxies (KS test).}
\label{fig:distri_gal}
\end{figure}

In Fig.~\ref{fig:distri_gal}, we show the PDF of galaxy mass for all galaxies in the simulated volume as grey distributions in the top and bottom panels. The distributions are normalized by the total number $N_{\rm tot}$ of galaxies in the respective samples; $N_{\rm tot}$ are given in Section 3.2. 
In the top panel, we also show the distribution of galaxies at the edges of the voids, with void-centric distances of $0.8\leqslant r/R_{\rm void}\leqslant 1$ (blue histogram), and the mass distribution of inner void galaxies ($r/R_{\rm void}\leqslant 0.65$, red histogram). 
For comparison, in the bottom panel we show the mass distribution of galaxies embedded in large Voronoi cell with a volume larger than $10\, \rm (cMpc/h)^3$ (blue histogram), and the distribution of even more isolated galaxies with $\geqslant 20 \rm (cMpc/h)^3$ (red histogram).
All the distributions are normalized to the total number of galaxies in the corresponding sample. Star symbols in both panels represent the mean of the distributions.

The median values $\tilde{M}_{\rm gal}=1.63,\, 2.13, \,2.23\times 10^{9}\, \rm M_{\odot}$ are found for the inner galaxies, the shell galaxies, and all the simulated galaxies, respectively. For the mean values\footnote{We also verified that the mean of $\log_{10} M_{\rm gal}$ also decreases with decreasing $r/R_{\rm void}$ with $\langle \log_{10}\, M_{\rm gal}/\rm M_{\odot}\rangle=9.27 \,{\rm (inner)},\, 9.39\,{\rm (shell)}, \, 9.43\,  {\rm (all)}$.}, we find $\langle M_{\rm gal}\rangle=5.90\times 10^{9} \, \rm M_{\odot}\,{\rm (inner)},\, 1.12 \times 10^{10}  \, \rm M_{\odot}\,{\rm (shell)}, \,1.27\times 10^{10}\, \rm M_{\odot}\,  {\rm (all)}$. We note here that while the median and the mean of the distributions for inner void galaxies are lower, the difference with shell galaxies and the entire galaxy population of the box are small, and the noise due to small number statistics high.

The mass distribution of the shell galaxies is indistinguishable from the distribution of all the galaxies in the box. However, with a KS test ($p=0.006$) we find that the distribution of the inner void galaxies is statistically different.
The relative abundance of low-mass galaxies (with $M_{\rm gal}\leqslant 10^{10}\, \rm M_{\odot}$) is higher in the inner region of voids than in the full simulation.

The selection of void galaxies with the volume of their Voronoi cell allows us to select galaxies without close neighbors, i.e., the most isolated galaxies. In the bottom panel of Fig.~\ref{fig:distri_gal}, we find that the distributions of isolated galaxies with $V_{\rm gal}>10 \, \rm (cMpc/h)^{3}$ and $V_{\rm gal}>20 \, \rm (cMpc/h)^{3}$ are both distinct from the distribution of all galaxies (KS test $p=2\times 10^{-5}, \, 2\times 10^{-4}$ respectively). 
The median values of the distributions are also lower than for all galaxies with $\tilde{M}_{\rm gal}=1.44, 1.28\times 10^{9}\, \rm M_{\odot}$ for $V_{\rm gal}>10\, {\rm and} \, 20\, \rm (cMpc/h)^{3}$, respectively. 
Isolated galaxies have preferentially low mass of $M_{\rm gal}\leqslant 10^{10}\, \rm M_{\odot}$.

\subsection{Stellar mass to halo mass relation}
\begin{figure}
\centering
\includegraphics[scale=0.52]{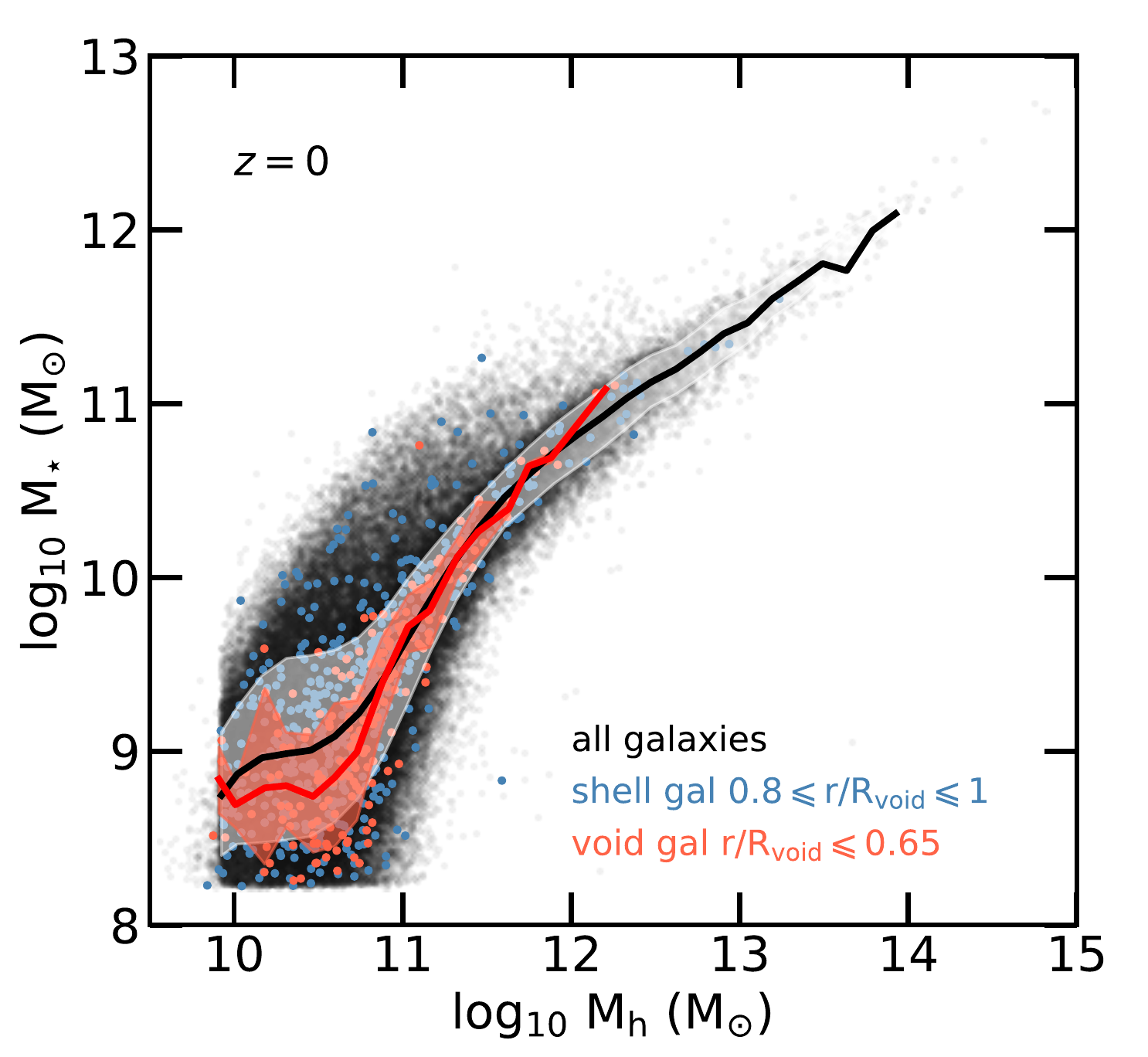}
\includegraphics[scale=0.52]{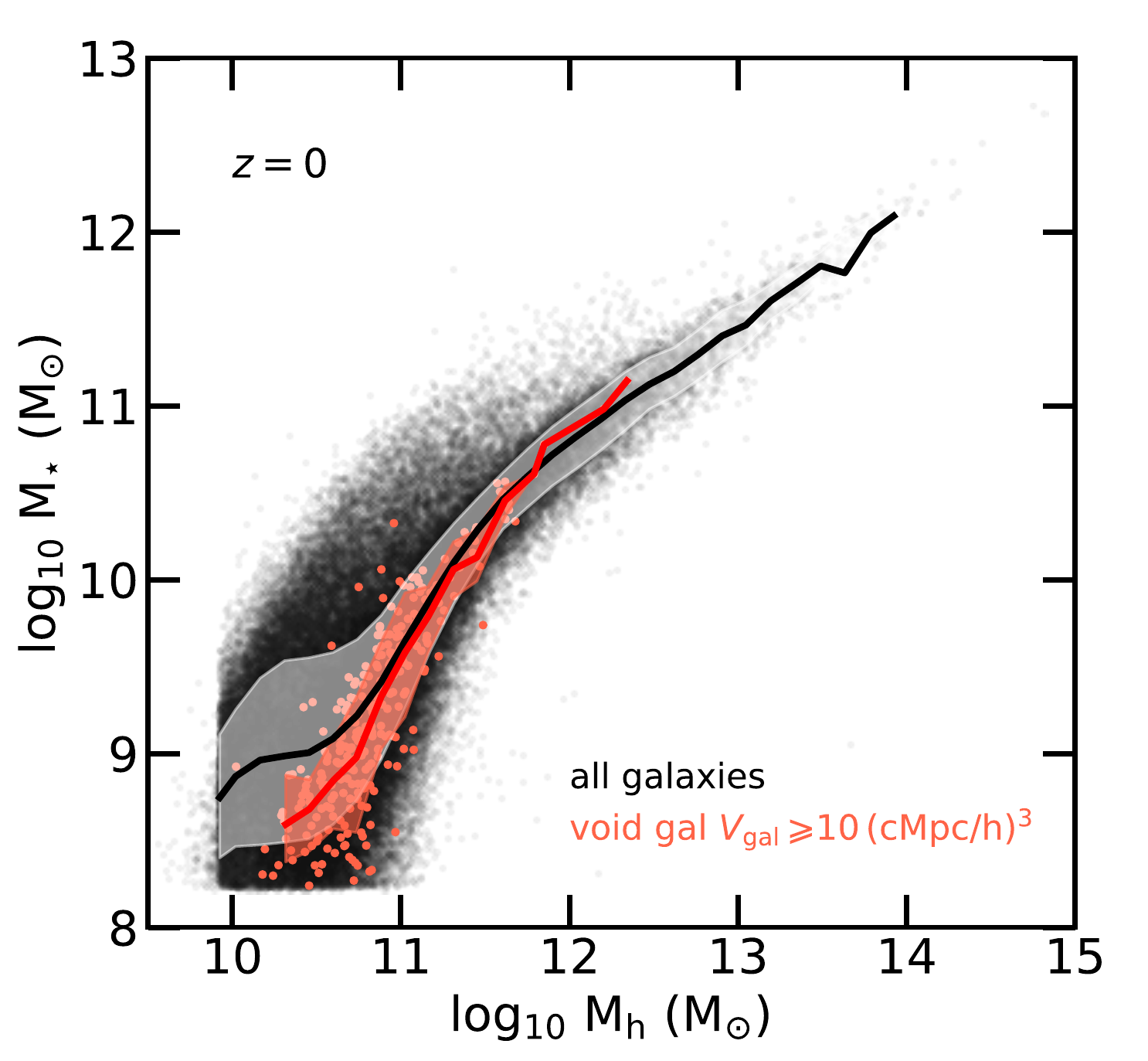}
\caption{{\it Top panel}: Stellar mass - halo mass relation for all galaxies in the simulation in black, galaxies located in the shells of voids in blue, and located in the inner regions of voids in red. Solid lines show the mean relation, and shaded areas the 15-85th percentiles of the distributions in fixed dark matter halo mass. The stellar mass distributions (at fixed mass halos) are distinct for $M_{\rm h}\leqslant 10^{10.75}\, \rm M_{\odot}$ (KS test). {\it Bottom panel:} The red line and dots show the void galaxies with large Voronoi cells. Their distribution (at fixed halo mass) is different from the distribution for the entire simulation up to $M_{\rm h}\leqslant 10^{11.25}\, \rm M_{\odot}$ (KS test). Isolated or inner void galaxies have lower stellar masses than what we could expect for their halo masses.}
\label{fig:mstar_mhalo}
\end{figure}

We show the relation between the stellar mass of galaxies and the mass of their host dark matter halos in black dots in Fig.~\ref{fig:mstar_mhalo} (top panel). The mean relation is shown as a black line and the 15th and 85th percentiles as the grey shaded area. We show in blue the shell galaxies and in red the galaxies that are located at the inner regions of voids.
With a KS test, we find that at fixed halo mass bins, the galaxy stellar mass distribution of the void inner regions is different from the distribution of all galaxies
up to $M_{\rm h}\leqslant 10^{10.75}\, \rm M_{\odot}$. On average, and at fixed dark matter halo mass, galaxies in the inner regions of voids have lower stellar masses in low-mass halos of $M_{\rm h}\leqslant 10^{10.75}\, \rm M_{\odot}$ than in the full simulation.
The mean relation for the shell galaxies (not shown here) is very similar to the mean relation for the entire simulation.

Similarly, the bottom panel of Fig.~\ref{fig:mstar_mhalo} shows the same relation, but we highlight in red the population of galaxies that are isolated in the Voronoi cell with $V_{\rm gal}\geqslant 10 \, \rm (cMpc/h)^{3}$. The evolution of these isolated void galaxies also seems different compared to the average evolution in the simulation box. With a KS test, we find that these galaxies have different stellar mass distributions, and have lower stellar mass compared to what we could expect for halos of $M_{\rm h}\leqslant 10^{11.25}\, \rm M_{\odot}$.

In the previous section, we have seen that the cosmic voids that we identified in Horizon-AGN preferentially host low-mass galaxies with $M_{\rm gal}\leqslant 10^{10}\, \rm M_{\odot}$, compared to what we could expect from the mass distribution of the entire simulated volume of the simulation. In this section, we have shown that for the isolated or the inner void galaxies in relatively low-mass halos with $M_{\rm h}\leqslant 10^{11}\, \rm M_{\odot}$ the galaxies were, on average, less massive than the other galaxies of the simulation embedded in same mass halos.

\subsection{Merger history of void galaxies} 
Because void galaxies are embedded in underdense regions of the Universe where few to no galaxy mergers are expected, we can expect galaxy evolution in voids to be driven mainly by {\it in-situ} processes. The in-situ stellar fraction of a galaxy is defined as the cumulative mass of stars formed in its main progenitors, and not in galaxies that have been merged during its lifetime.

We derive the percentages of galaxies with an in-situ fraction higher than $f_{\rm in-situ}=70\%$ in three different mass bins: $10^{8}\leqslant M_{\rm gal}< 10^{9}\, \rm M_{\odot}$, $10^{9}\leqslant M_{\rm gal}< 10^{10}\, \rm M_{\odot}$, and $10^{10}\leqslant M_{\rm gal}< 10^{11}\, \rm M_{\odot}$. 
Our threshold of $f_{\rm in-situ}$ is arbitrarily chosen, and quite high since most of the galaxies studied here have low masses and therefore are expected to have high in-situ fractions. 
Indeed, we find that $98\%$, $97\%$, of all the simulated galaxies in the stellar mass bins $10^{8}\leqslant M_{\rm gal}< 10^{9}\, \rm M_{\odot}$ and $10^{9}\leqslant M_{\rm gal}< 10^{10}\, \rm M_{\odot}$, respectively, have more than $70\%$ of their stellar content formed in-situ. Here, we confirm that this is also the case for void galaxies, where even a higher percentage of inner void galaxies have high in-situ fractions ($98.5\%$ and $98\%$ for the two first bins). 
We find that the percentages of galaxies with high in-situ fractions are lower for galaxies located in the void shells: $97.2\%$ and $96.4\%$ for the two first galaxy mass bins, respectively. 
This indicates that shell galaxies (even with quite low masses $\leqslant 10^{9}\, \rm M_{\odot}$) are more likely to have had mixed stellar formation (in-situ and ex-situ).
We find similar results for isolated galaxies, and note that all the most isolated galaxies with $V_{\rm gal}\geqslant 20 \, \rm (cMpc/h)^{3}$ have higher in-situ fractions than $f_{\rm in-situ}\geqslant 70\%$. 
These conclusions are unchanged for the most massive bin $10^{10}\leqslant M_{\rm gal}< 10^{11}\, \rm M_{\odot}$, but we do find larger ex-situ stellar contents in these more massive galaxies (e.g., $62.4\%$ of all galaxies in the simulation have in-situ fractions higher than $70\%$).

Here we have confirmed that, as expected, void galaxies in the inner regions of voids or isolated galaxies in void do not have large ex-situ stellar formation in their lifetime. We note that the percentages of void galaxies with large in-situ fractions is, on average, not so different from the average over all the galaxies in the simulation, and this because we focus on small voids here and so quite low-mass galaxies.
We find that selecting isolated galaxies with large Voronoi cell volume can be efficient to build a sample of galaxies with very large in-situ fractions. The larger the cell volume, the more efficient the selection of in-situ stellar formation galaxies. We find that galaxies located in voids are, on average, mainly driven by in-situ processes.


\subsection{Star formation activity of void galaxies} 
\subsubsection{Star formation rate of inner void and shell galaxies}

\begin{figure*}
\centering
\includegraphics[scale=0.66]{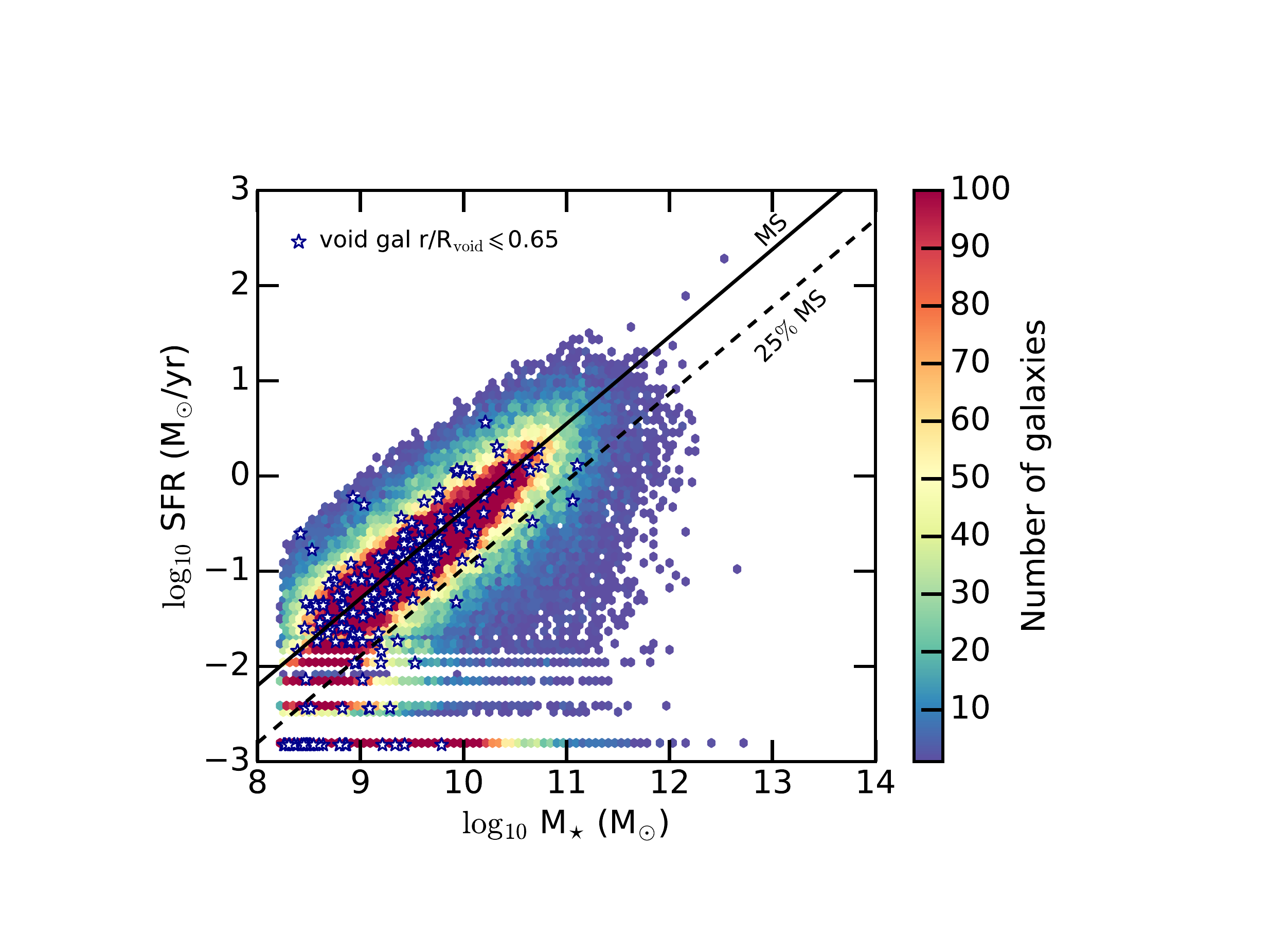}
\caption{Star formation rate of the galaxies in Horizon-AGN as a function of their total stellar mass, at $z=0$. The diagram is divided in hexabins color-coded by the number of galaxies in each bin. The solid black line indicates the star-forming sequence (MS) of the galaxy population and the dashed line our limit to define star-forming vs quiescent galaxies. Inner void galaxies with $r/R_{\rm void}\leqslant 0.65$ are shown as blue stars. A large fraction of them are still around the star-forming sequence, and therefore forming stars efficiently.}
\label{fig:distri_sfr}
\end{figure*}

In Fig.~\ref{fig:distri_sfr} we show how the simulated galaxies are distributed in the SFR - $M_{\rm gal}$ plane\footnote{The stripes at fixed SFR in Fig.~\ref{fig:distri_sfr} represent galaxies with only a few stellar particles.}.
Star formation rates are averaged over 1 Gyr. For visual purposes, we set a lower limit of $\log_{10}\rm SFR/(M_{\odot}/yr)= -3$ (shown at $\log_{10}\rm SFR/(M_{\odot}/yr)= -2.8$ in Fig.~\ref{fig:distri_sfr}) for galaxies with a lower star formation rate, including those that are not forming stars (to display them on the figure). The diagram is divided in hexabins color-coded by the number of galaxies in each bin. First, we define the star-forming sequence (also called main-sequence, MS hereafter) of galaxies, by taking the mean SFR in the galaxy mass range $M_{\rm gal}=10^{9}-10^{10}\, \rm M_{\odot}$, for which we do not expect to find a large fraction of quenched galaxies (i.e., galaxies with reduced SFR, or null SFR). The star-forming sequence is defined by:
\begin{eqnarray}
\log_{10} \rm{SFR_{\rm MS}}=\alpha + \beta \log_{10}\left( \frac{M_{\rm gal}}{10^{10}}\right),
\end{eqnarray}
with $\alpha=-0.368$ and $\beta=0.917$. We use this relation for the entire galaxy mass range studied here.
We show the star-forming sequence as a solid black line. Our limit to define star-forming galaxies and quiescent galaxies is $25\%$ below the main star-forming sequence \citep[e.g.,][]{2015MNRAS.451.2933B,2019MNRAS.484.4413H}, and is shown as a dashed black line in Fig.~\ref{fig:distri_sfr}. 
Galaxies with $\rm SFR \geqslant 25\% \, SFR_{MS}$ are considered as star-forming galaxies. Quiescent galaxies are galaxies with reduced or null star formation rate, sustained in time. Here we define quiescent galaxies as galaxies with $\rm SFR < 25\% \, SFR_{MS}$.

Blue star symbols indicate the 195 void galaxies with $r/R_{\rm void}\leqslant 0.65$. Most of these galaxies are forming stars. Only $13\%$ of these inner void galaxies with $M_{\rm gal}\geqslant 10^{9}\, \rm M_{\odot}$ are quiescent galaxies, $24\%$ in total if we include galaxies with lower stellar mass. Even if a large fraction of inner void galaxies are on the star-forming sequence, we find with a KS test that the SFR distributions of these galaxies, binned in 0.25 dex of stellar mass, are consistent with the distributions for the entire simulation. Only the distribution of SFR for the mass bin $\log_{10} M_{\rm gal}/\rm M_{\odot}=8.5-8.75$ is different. We do not show the distributions here. However, our findings will need to be checked with larger samples of galaxies as each of the bins here only include less than 30 galaxies.  The median SFR value in these bins is always similar or higher for the inner void galaxy distributions than the full simulation distributions. 
For clarity, we do not show the location of shell galaxies with $0.8 \leqslant r/R_{\rm void}\leqslant 1$, but they populate a much broader part of the SFR-$M_{\rm gal}$ diagram. The fraction of quiescent galaxies is higher in shell galaxies, with $27\%$ among the $M_{\rm gal}\geqslant 10^{9}\, \rm M_{\odot}$, and $37\%$ independently of galaxy stellar mass. 

We look at the SFR - $M_{\rm gal}$ plane of isolated galaxies with $V_{\rm gal}> 10\, \rm (cMpc/h)^{3}$ (not shown here). We find that the median SFR values of the distributions, binned in 0.25 dex of stellar mass, are always higher than for the full simulation distribution. While we could not reject a null hypothesis when comparing the SFR distributions in stellar mass bins for the inner void galaxies and the full simulation distributions, we find with a KS test that the distributions of isolated galaxies are distinct for $M_{\rm gal}\leqslant 10^{9.5}\, \rm M_{\odot}$. 

As a conclusion, we find here that a high fraction of inner void galaxies are still active, i.e. qualified as star-forming galaxies. 
To define star-forming galaxies and quiescent galaxies in this paper, we have used a cut of $25\%$ dex of the star-forming sequence. With this conservative choice, we find that we can not rule out that the SFR distribution of inner galaxies ($r/R_{\rm void}\leqslant 0.65$) is statistically different from the full simulation, but we can for isolated galaxies $V_{\rm gal}> 10\, \rm (cMpc/h)^{3}$.
If we would instead define quiescent galaxies as having null star-formation rate, the fraction of star-forming galaxies would be higher. For example, the star-forming fraction of inner void galaxies would be equal to unity for $M_{\rm star}\geqslant 10^{10}\, \rm M_{\odot}$.



\subsubsection{Star formation properties as a function of void-centric distances}

\begin{figure*}
\centering
\includegraphics[scale=0.5]{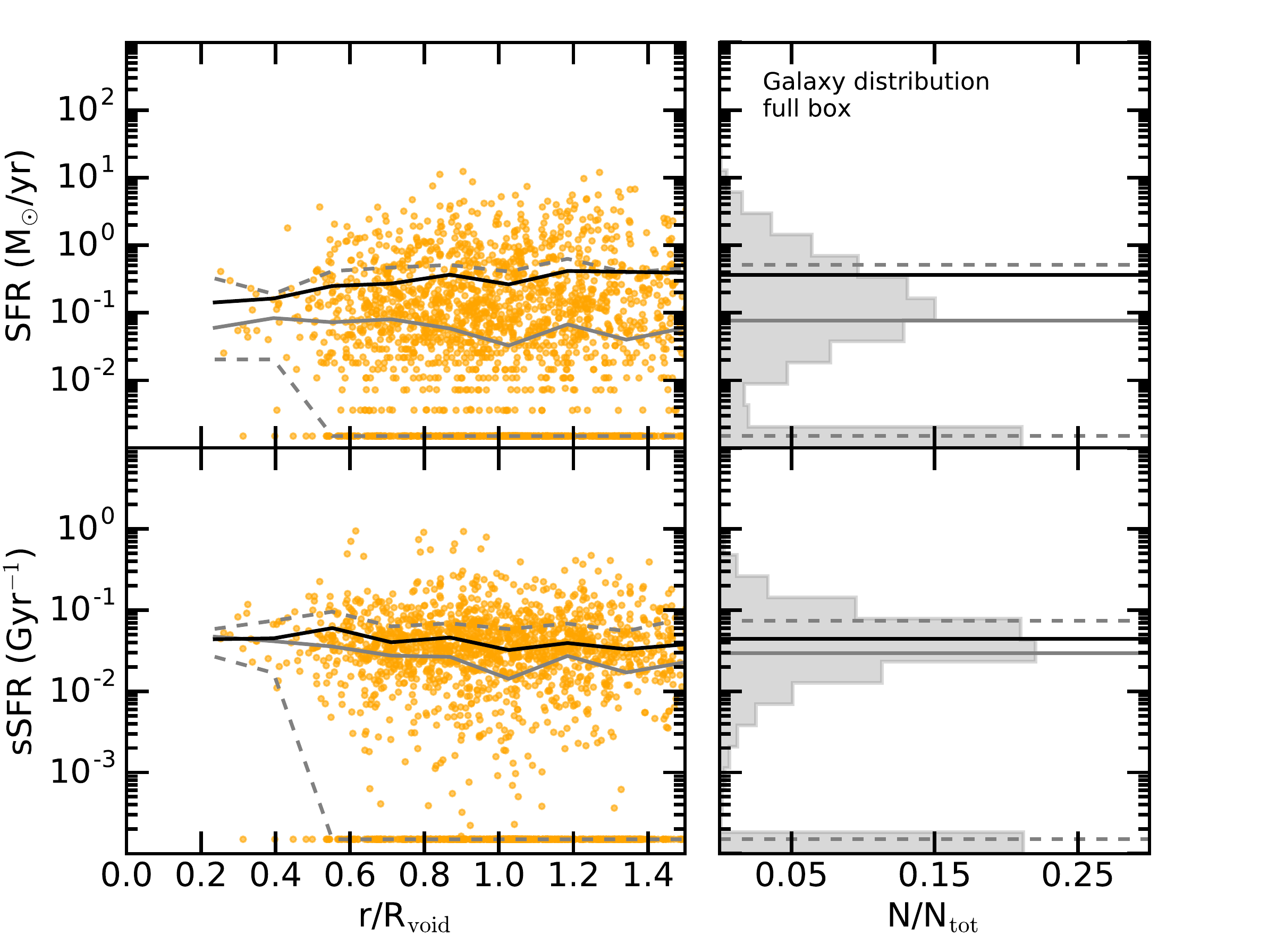}
\caption{Galaxy star formation rate SFR (top panel) and specific star formation rate sSFR (bottom panel) as a function of normalized void-centric radius $r/R_{\rm void}$. Grey distribution in the right panel shows the galaxy mass distribution for the full simulation box. On both panels, the black lines indicate the mean of the distributions, the solid grey lines the median of the distributions, and the dashed grey lines the $\pm \sigma$. The mean SFR of void galaxies decreases when going further into the void, but the sSFR slightly increases. This means that at least some inner void galaxies form stars slightly more efficiently with respect to their masses, even if they can have lower SFR due to their low masses.}
\label{fig:distri_sfr_rad}
\end{figure*}

We now turn to study the star-forming properties of void galaxies as a function of their void-centric distances. We show the SFR of galaxies in Fig.~\ref{fig:distri_sfr_rad} (top panel). The distribution of the SFR for all simulated galaxies is shown on the right panel for comparison. As previously, the mean value binned in $r/R_{\rm void}$ is shown with the black line on the left panel; we also show the mean value of the total distribution in black in the right panel. Similarly, solid grey lines indicate the median of the distributions, and the dashed grey lines the $\pm \sigma$ of the distributions.
The mean SFR decreases when approaching the center of cosmic voids from a mean value of $\rm \langle SFR\rangle \sim 0.4\, M_{\odot}/yr$ at $r/R_{\rm void}=1.5$, which is similar to the mean of the SFR distribution of the simulation ($\rm \langle SFR\rangle \sim 0.36\, M_{\odot}/yr$), to $\rm \langle SFR\rangle \sim 0.25 \, M_{\odot}/yr$ at $r/R_{\rm void}=0.5$. We report these values in Table~\ref{table:void_centric}.
In the bottom panels of Fig.~\ref{fig:distri_sfr_rad}, we also show the specific star formation rate sSFR of the void galaxies and the distribution for the entire simulation. 
While the stellar mass and the SFR can be significantly biased by the relative abundance of low-mass galaxies in voids, the sSFR is less impacted.
The mean sSFR slightly increases for void galaxies with smaller void-centric radii. It demonstrates that the SFR of void galaxies is slightly lower for inner galaxies on average, but they could be forming stars slightly more efficiently with respect to their galaxy stellar mass than galaxies in the rest of the simulated volume.

\begin{table*}
\caption{Mean values of the galaxy total stellar mass of void galaxies ($M_{\rm gal}$), star formation rate (SFR), specific star formation rate (sSFR), and BH mass ($M_{\rm BH}$), for void-centric distances of $r/R_{\rm void}\sim 0.5,1,1.5$. For comparison, we also report the mean value of all galaxies present in the simulation on the last right column.}
\begin{center}
\begin{tabular}{ccccc}
\hline
& \multicolumn{3}{c}{Void galaxies} & All galaxies\\
& $r/R_{\rm void}\sim 0.5$ & $r/R_{\rm void}\sim 1.0$ & $r/R_{\rm void}\sim 1.5$ & \\  
\hline
$<M_{\rm gal}> \,\rm (M_{\odot})\, \,$ & $6.2\times 10^{9}$ & $1.4\times 10^{10}$ & $1.3\times 10^{10}$ & $1.3\times 10^{10}$\\
\hline
$\rm <SFR>\, (M_{\odot}/yr)$ & 0.25 & 0.35 & 0.40 & 0.36  \\
\hline
$\rm <sSFR> \, (Gyr^{-1})$ & 0.060 & 0.039 & 0.038 & 0.044 \\
\hline
$<M_{\rm BH}> \, (\rm M_{\odot})$ & $2.3 \times 10^{7}$ & $8.4\times 10^{7}$ & $5.8\times 10^{7}$ & $5.5 \times 10^{7}$\\ 
\hline
\end{tabular}
\end{center}
\label{table:void_centric}
\end{table*}

\subsubsection{Fractions of star-forming galaxies as a function of void-centric distances}
\begin{figure}
\centering
\includegraphics[scale=0.55]{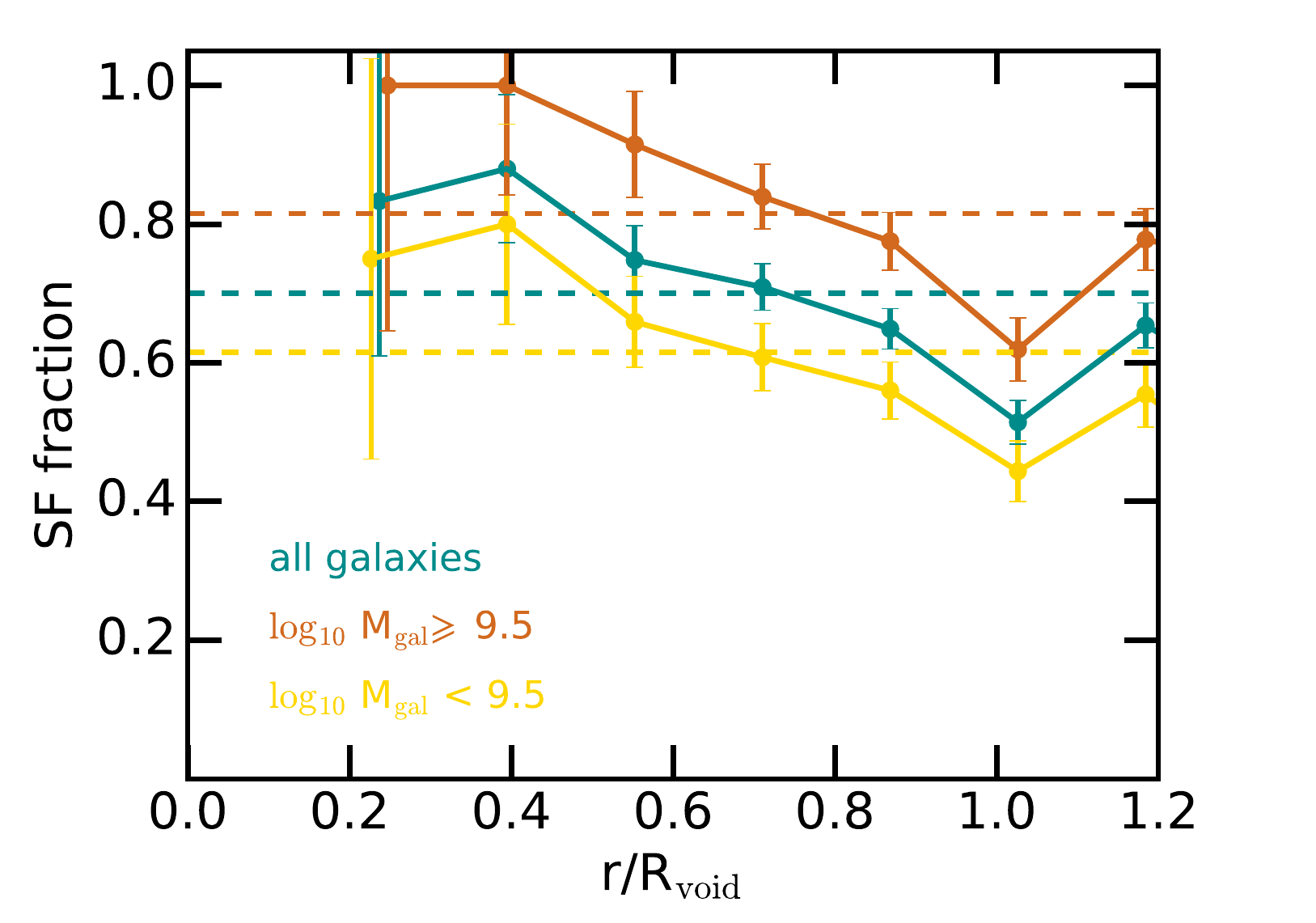}
\caption{Fraction of star-forming galaxies as a function of normalized void-centric radius for all galaxies (green solid line), low-mass galaxies with $M_{\rm gal}<10^{9.5}\, \rm M_{\odot}$ (yellow solid line), and higher mass galaxies with $M_{\rm gal}\geqslant 10^{9.5}\, \rm M_{\odot}$ (brown solid line). Error bars are Poissonian (calculated using the number of galaxies in each radius bin). To guide the eyes, we also show the star-forming fractions averaged over all galaxies in the simulation box, as dashed horizontal lines.}
\label{fig:fraction_sf}
\end{figure}

We quantify the fraction of void galaxies that are star-forming and compare it to the full box galaxy population. 
Star-forming galaxies are defined as having SFR (averaged over 1 Gyr) higher than $25\%$ of the main star-forming sequence. 
We first report the fraction of galaxies in the entire simulation box that are forming stars as a green dashed line in Fig.~\ref{fig:fraction_sf}. Similarly, we show the star-forming fraction of two selections of galaxy stellar mass: $M_{\rm gal}<10^{9.5}\, \rm M_{\odot}$ as a dashed yellow line, and $M_{\rm gal}\geqslant 10^{9.5}\, \rm M_{\odot}$ as a dashed brown line. We compare these fractions to the fraction of star-forming galaxies in cosmic voids, shown as solid lines in Fig.~\ref{fig:fraction_sf}. We show Poissonian error bars (calculated using the number of galaxies in each radius bin). The fraction of star-forming galaxies is higher in the inner regions of voids than in the full simulation.
All void fractions are higher than average for $r/R_{\rm void}\leqslant 0.7$, and decrease from the center of voids to $r/R_{\rm void}=1$, before oscillating around a constant value. 
This constant value appears to be lower than the average over all the simulated galaxies of the box (dashed lines) because only including galaxies linked to a given void by the void finder algorithm. 
We also note that the fraction of star-forming galaxies among the $M_{\rm gal}\geqslant 10^{9.5}\, \rm M_{\odot}$ galaxies (brown lines) is higher than the average for all galaxies (green lines). However, we find lower star-forming fractions as we use higher stellar mass limit of $M_{\rm gal}\geqslant 10^{10.5}\, \rm M_{\odot}$ (not shown in Fig.~\ref{fig:fraction_sf} for clarity), for which the fraction of quenched galaxies starts increasing. 

Dividing the sample of void galaxies into low- and high-mass galaxies in Fig.~\ref{fig:fraction_sf} allows us to say that the abundance of star-forming galaxies in the inner region of voids is not only due to the prevalence of low-mass galaxies (which are generally predicted to be star-forming). Indeed, we find the same trend for different samples of galaxy stellar mass. 

The mass bias in the void galaxy properties, due to the relative abundance of low-mass galaxies, is still subject to intense debate in the literature \citep[in observations,][and references therein]{ 2014MNRAS.445.4045R,2017ApJ...846L...4R}. We investigate this in more detail for our sample in the following section. 

\begin{figure}
\centering
\includegraphics[scale=0.5]{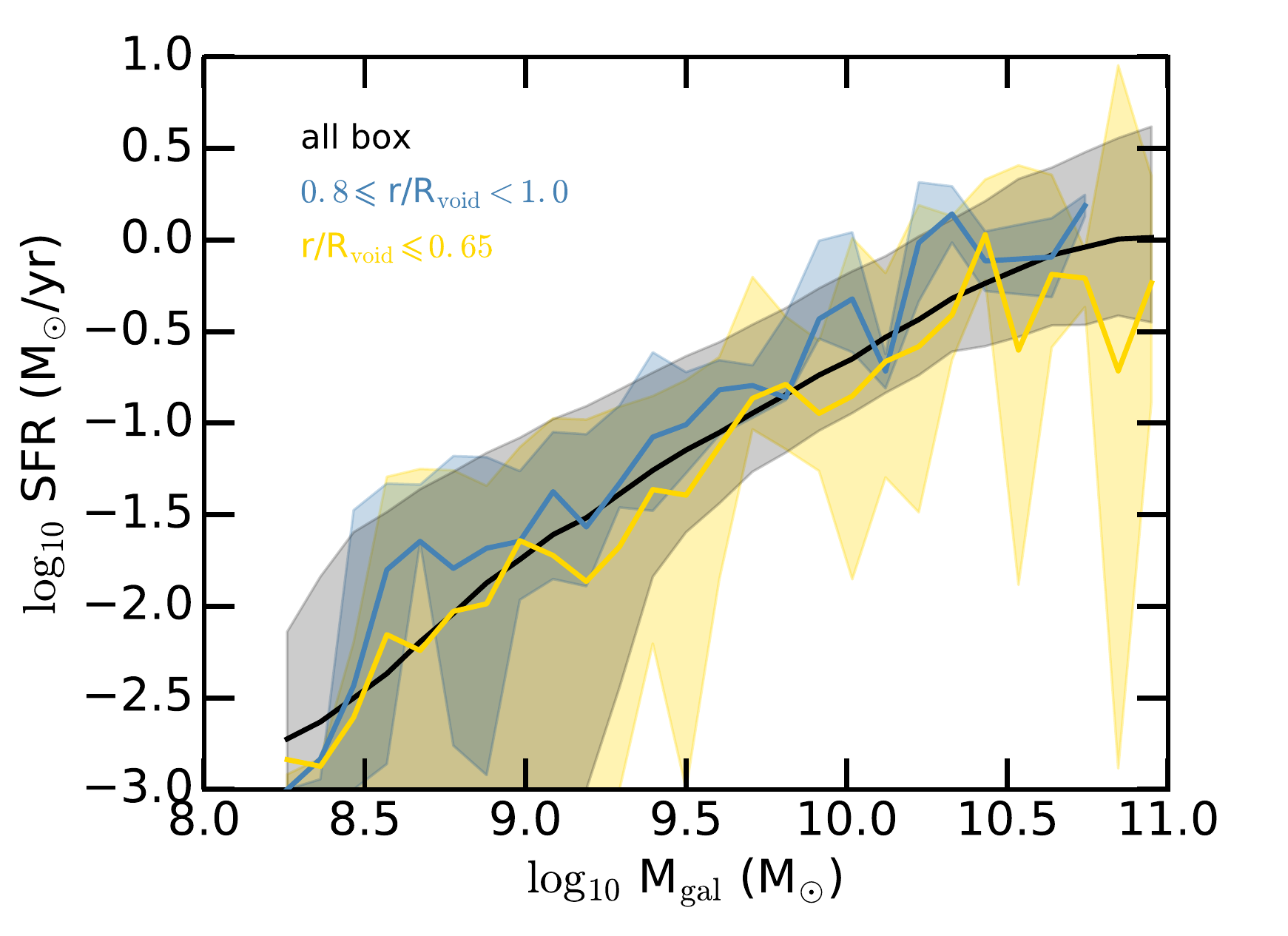}
\caption{Mean star formation rate as a function of galaxy mass considering all galaxies in the simulation (black solid line), galaxies in walls defined as $\rm 0.8 \leqslant r/R_{void} < 1.0$ (blue solid line), and inner void galaxies defined as $\rm r/R_{void} \geqslant 0.8$ (yellow solid line).
Shaded areas show the standard deviation $\pm \sigma$ of the distributions. At a given galaxy mass, galaxies embedded in the central region of voids have, on average, lower SFR than galaxies of the same mass in walls. This is independent of galaxy stellar mass, which shows that there is no strong mass bias.}
\label{fig:biased_gal}
\end{figure}

\subsection{Impact of the abundance of low-mass galaxies in voids?}
To investigate the bias toward the abundance of low-mass galaxies in our results, we show in Fig.~\ref{fig:biased_gal} the mean star formation rate as a function of galaxy stellar mass, for all galaxies in the simulation (black line), for shell galaxies ($0.8\leqslant r/R_{\rm void}<1.0$, blue line), and for inner galaxies ($r/R_{\rm void}\leqslant 0.65$, yellow line). Shaded regions represent the standard deviation $\pm \sigma$ of the samples. Here, we set the minimum value $\log_{10} \rm SFR=10^{-3}\, \rm M_{\odot}/yr$ to the galaxies with null star formation rate. 

Shell galaxies tend to have a mean SFR higher than galaxies located in the inner regions of voids. This effect stands for all galaxy mass bins. We also find that the mean SFR of the two respective populations lies within the standard deviation of the other distribution, meaning that there is a large variance of possible SFR among the galaxies of the two samples.
Moreover, shell galaxies have higher SFR than the mean of the entire distribution of simulated galaxies, whereas inner void galaxies are, on average, below the mean relation for all simulated galaxies (black line in Fig.~\ref{fig:biased_gal}). 

From Fig.~\ref{fig:distri_sfr_rad}, we find that the SFR drops when approaching the inner regions of voids. This effect could have been biased due to the high relative abundance of low-mass galaxies in voids (see Fig.~\ref{fig:distri_gal}). However, here we find that for all galaxy mass bins the SFR in voids is lower than the SFR in void shell galaxies. We conclude that in this simulation, the lower SFR in the inner regions of voids can not be interpreted only as a consequence of the low-mass galaxy bias.
However, the effect we find here is mild and needs to be investigated in more detail using larger samples of galaxies but also larger simulated or observed volume to further push these results to better statistics.


\section{Properties of supermassive black holes embedded in cosmic voids}

We now turn to investigate the properties of BHs in void galaxies. In this paper, we particularly focus on the BH mass distribution, their mass ratio with the host galaxy total stellar mass, their formation time, and their ability to accrete gas.
We discuss the BH galaxy occupation fraction in the discussion section at the end of the paper.

BH formation and growth in low-density regions of the Universe have been studied by means of zoom-in hydrodynamical cosmological simulations, i.e., by performing a simulation of a smaller region of the initial simulated volume but with a much higher resolution.
Using a semi-analytical model\footnote{The semi-analytical model of \citet{2015ApJ...799..178K} assumes that the properties of BHs scale with the properties of their host galaxies.} on top of the merger-trees extracted from zoom-in simulations, \citet{2015ApJ...799..178K} find that in void regions the contribution of BH-BH mergers on BH growth can be as low as $1\%$, against $2.5 \%$ in cluster regions. In their work, gas accretion was the dominant process for BH growth. 
Instead of analyzing a few zoom-in void regions, here we investigate BH evolution {\it self-consistently} and {\it statistically} in the cosmic voids of the large-scale cosmological simulation Horizon-AGN.

\subsection{BH mass properties}
\begin{figure}
\centering
\includegraphics[scale=0.355]{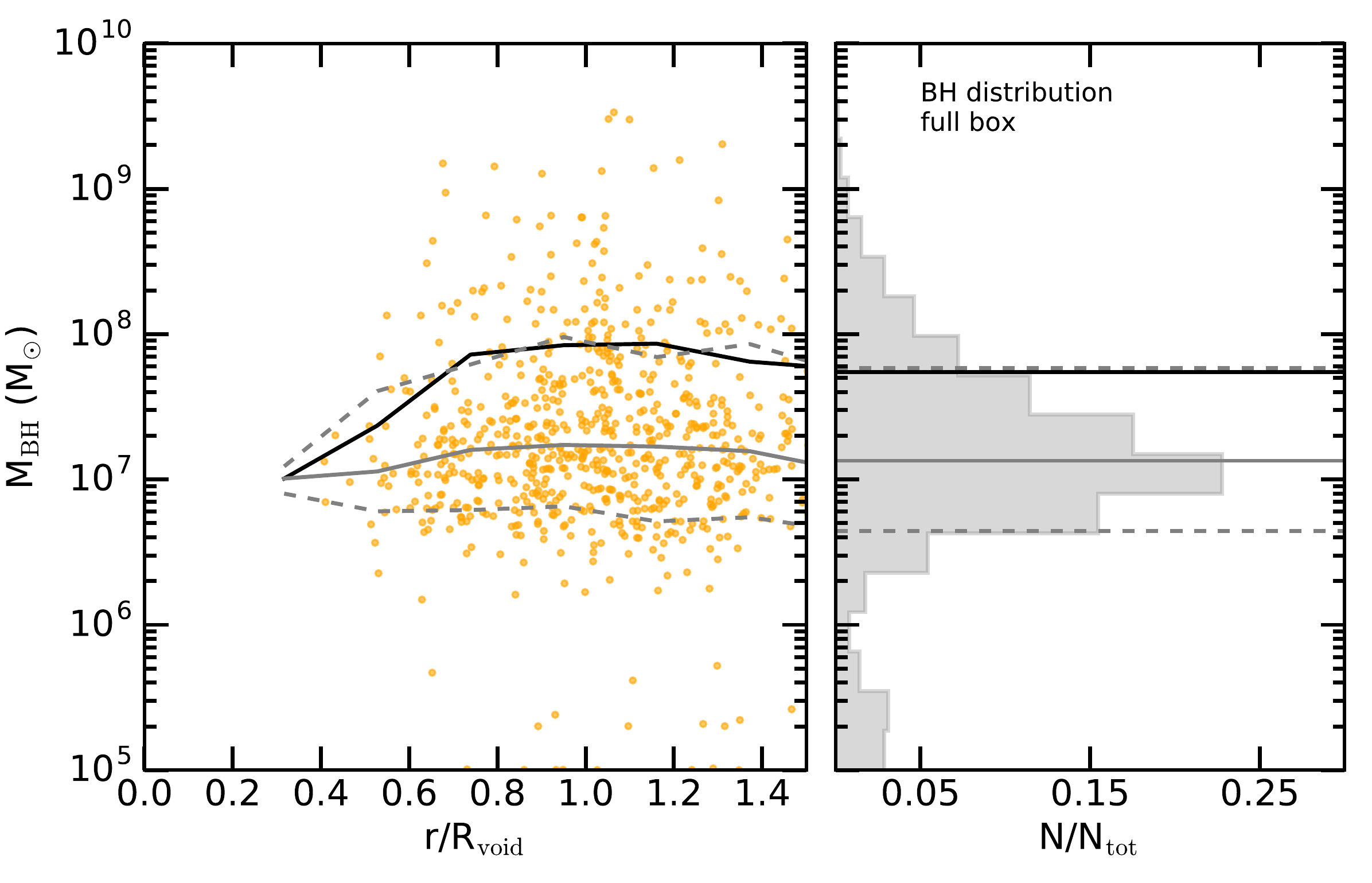}
\caption{BH mass as a function of normalized void-centric radius. Grey distribution in the right panel shows the galaxy mass distribution for the full simulation box. On both panels, the black lines indicate the mean of the distributions, the solid grey lines the median of the distributions, and the dashed grey lines the $\pm \sigma$. BHs are on average less massive (below the mean of the distribution of the entire simulated box) for lower and lower void-centric.}
\label{fig:distri_bh}
\end{figure}

We first study the mass distribution of BHs present in galaxies embedded in cosmic voids. Fig.~\ref{fig:distri_bh} shows the mass of BHs as a function of their void-centric distances. The black solid line shows the mean of the BH mass distribution, the grey solid and dashed lines represent its median and the standard deviation $\pm \sigma $. For comparison, we also show the mass distribution of all BHs formed in the simulation on the right panel, with the same definitions for the black and grey lines. We find that the relative abundance of lower-mass BHs increases when located closer to their void centers. In other words, the number of more massive BHs decreases with decreasing $r/R_{\rm void}$. This is similar to our findings for galaxy mass in Fig.~\ref{fig:distri_gal}, i.e., the prevalence of low-mass galaxies when moving to small void-centric distances. 
For both galaxies and BHs, we observe that the decline of the mean, and the small decline of the median, of the mass distribution takes place at about $r/R_{\rm void}\leqslant 0.8$.

With a KS test, we find that the mass distribution of the inner void BHs ($r/R_{\rm void}\leqslant 0.8$), as well as the distribution of the shell BHs ($0.8<r/R_{\rm void}\leqslant 1$), are statistically different from the distribution of the entire simulation ($p=0.087, 0.015$, respectively). However, we can not reject that the mass distribution of the inner BHs is statistically different from the distribution of shells BHs.

\subsection{Co-evolution between BHs and their host galaxies}
Both the means of BH and galaxy mass drop when approaching the center of voids. Can the presence of lower mass void BHs be interpreted as a consequence of the high number of low-mass galaxies in the inner regions of cosmic voids? To investigate this possible effect, we quantify how galaxies and their central BHs co-evolve together in the following.

\begin{figure}
\centering
\includegraphics[scale=0.5]{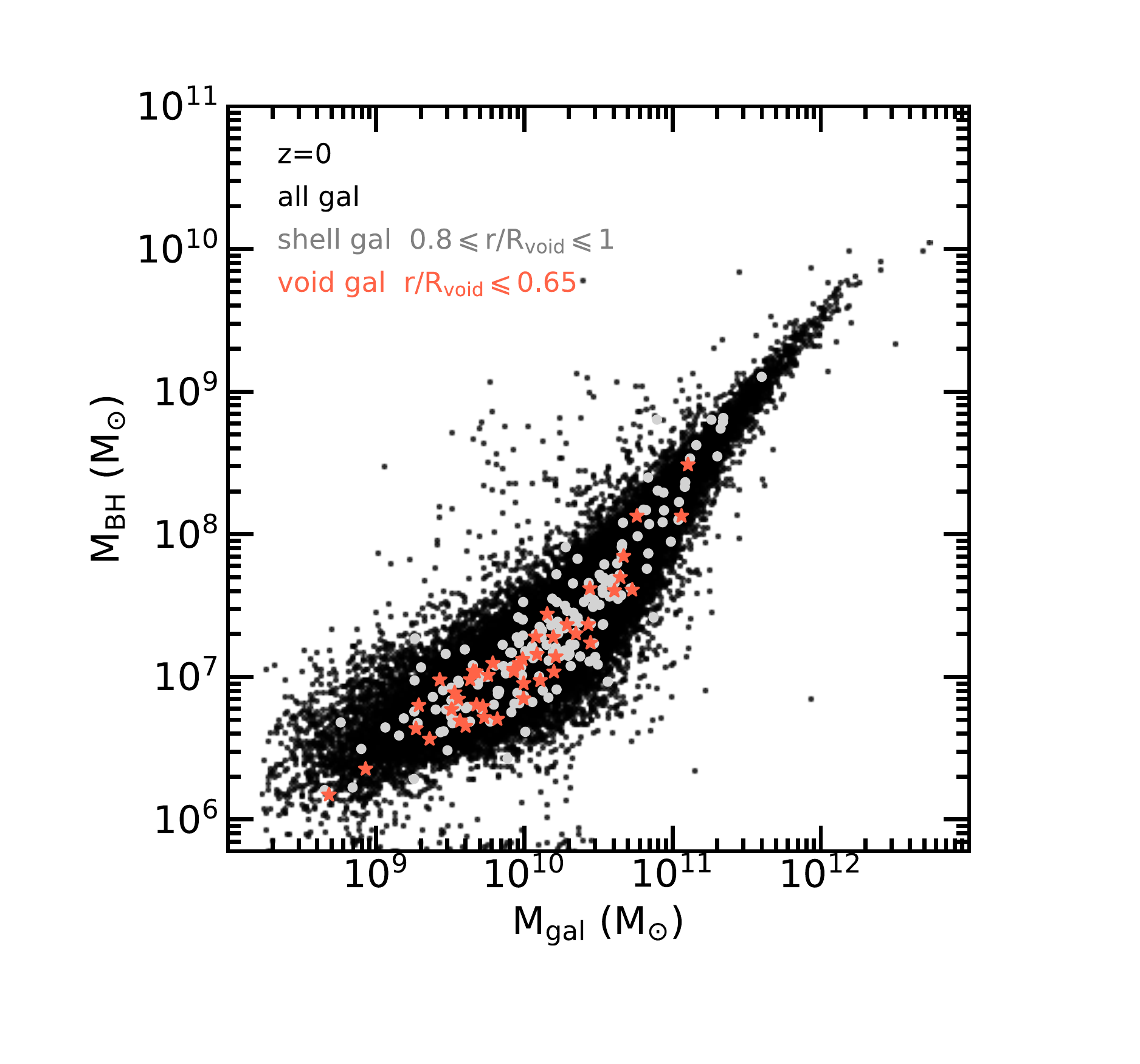}
\caption{BH mass - galaxy mass relation for all simulated galaxies in black, for shell galaxies with $0.8\leqslant r/R_{\rm void} \leqslant 1$ in grey, and void inner galaxies with void-centric radius $r/R_{\rm void}\leqslant 0.65$. BHs located in inner void galaxies or shell galaxies seem to follow the global behavior of the entire simulation.}
\label{fig:bhmass_galmass}
\end{figure}

\begin{table*}
\caption{Median and $\pm 1 \sigma$ of BH mass to total stellar mass ratio distribution for the three selections of galaxies showed in Fig.~\ref{fig:mass_ratio}. We find very similar parameters of the distributions for all the galaxies of the simulation and the inner void galaxies ($r/R_{\rm void}\leqslant 0.65$). We also report the parameters for different cuts, including both void-centric distance selection and Voronoi volume selection, in the last two rows. }
\begin{center}
\begin{tabular}{ccccccc}
\hline
& \multicolumn{2}{c}{all galaxies} & \multicolumn{2}{c}{$5\times 10^{8}\leqslant M_{\rm gal}< 10^{10}\, \rm M_{\odot}$} & \multicolumn{2}{c}{$M_{\rm gal}\geqslant 10^{11}\, \rm M_{\odot}$}\\
\hline
& median & $\pm 1\sigma$ & median & $\pm 1\sigma$ & median & $\pm 1\sigma$\\
\hline
all galaxies         & $-2.86$ & $-3.09,-2.61$ & $-2.77$ & $-3.00,-2.52$ & $-2.91$ & $-3.12,-2.69$ \\
void: $r/R_{\rm void}\leqslant 0.65$  & $-2.87$ & $-3.06,-2.63$ & $-2.80$ & $-2.96,-2.63$ & $-2.93$ & $-3.12,-2.78$ \\
void: $r/R_{\rm void}\leqslant 0.80, V_{\rm gal}>10\, \rm (cMpc/h)^{3}$ & $-2.87$ & $-3.04,-2.71$ & $-2.87$ & $-2.90,-2.65$ & $-3.03$ & $-3.17,-2.84$ \\  
void: $r/R_{\rm void}\leqslant 0.65, V_{\rm gal}>10 \,\rm (cMpc/h)^{3}$ & $-2.87$ & $-3.02,-2.69$ & $-2.79$ & $-2.87,-2.65$ & $-3.00$ & $-3.05,-2.78$ \\ 
\hline
\end{tabular}
\end{center}
\label{table:table_mass_ratio}
\end{table*}

\begin{figure*}
\centering
\includegraphics[scale=0.65]{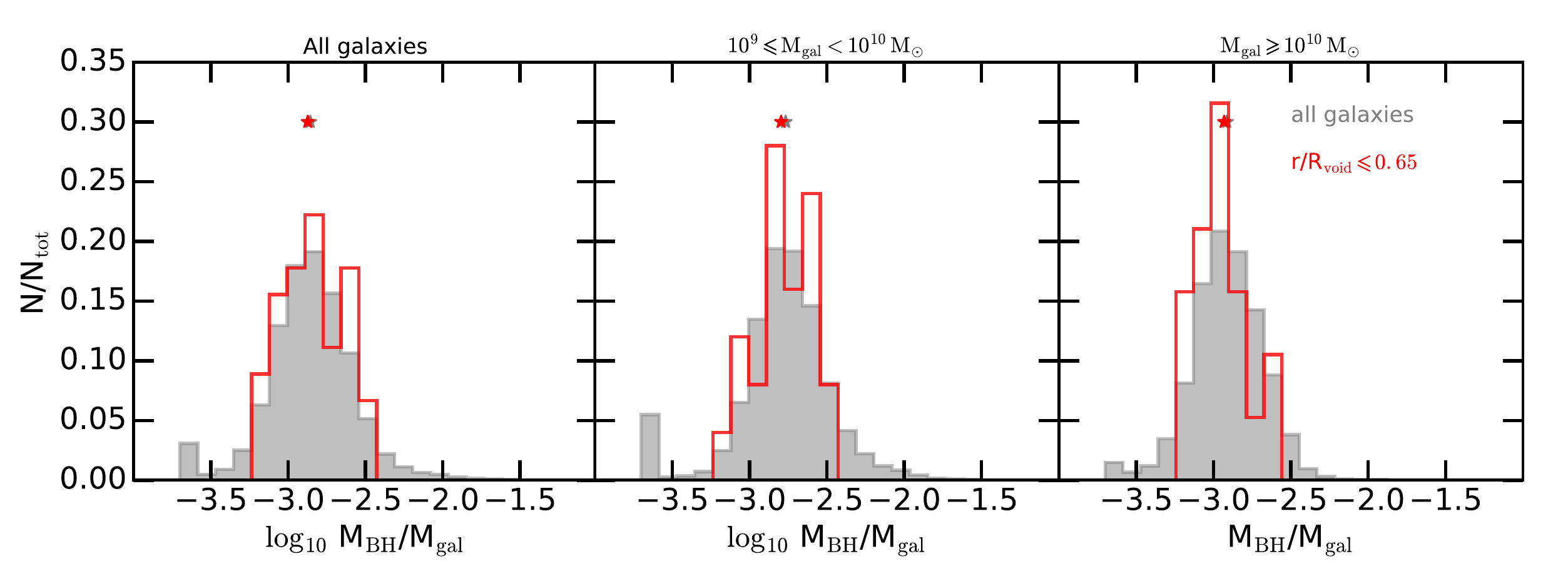}
\caption{Distributions of mass ratio between BHs and their host galaxies, for all simulated galaxies (in grey), or void galaxies with $\rm r/R_{void}\leqslant 0.65$ (in red), at $z=0$. Histograms are normalized by the number of elements in the samples. The left panel shows the distribution for galaxies independently of their mass, middle panel galaxies with $10^{9}\leqslant M_{\rm gal}< 10^{10} \, \rm M_{\odot}$, and right panel the most massive galaxies with $M_{\rm gal}\geqslant 10^{10}\, \rm M_{\odot}$. Star symbols indicate the median of the distributions. The distributions of mass ratios in inner void galaxies are very similar to the distributions for galaxies in the entire simulation, meaning that cosmic voids do not host under- or over-massive BHs compared to the mass of their host galaxies.}
\label{fig:mass_ratio}
\end{figure*}

\subsubsection{BH mass - galaxy stellar mass diagram}
We first present the BH mass - galaxy total stellar mass diagram in Fig.~\ref{fig:bhmass_galmass} at $z=0$. We show all the galaxies in the simulation as black dots. As in all cosmological hydrodynamical simulations, there is a tight correlation between BH mass and the total stellar mass of their host galaxies \citep[e.g., ][and Habouzit et al. (in prep) for a comparison of the scaling relation of 6 large-scale cosmological simulations]{2015MNRAS.452..575S,2016MNRAS.460.2979V}.
We represent the void shell galaxies as grey dots ($0.8 \leqslant r/R_{\rm void}\leqslant 1$), and the inner void galaxies ($r/R_{\rm void}\leqslant 0.65$) as red star symbols. Please note that here we do not show the BHs located within $0.65\leqslant r/R_{\rm void}\leqslant 0.8$. Interestingly, we see in Fig.~\ref{fig:bhmass_galmass} that our void sample do not include all the smallest galaxies and BHs. 
It is not surprising since we have made conservative choices when building our sample of void galaxies, our voids also have small sizes. Some of the galaxies on the lower left side of Fig.~\ref{fig:bhmass_galmass} could therefore turned out to be embedded on voids, but not included in our sample. A given fraction of these galaxies could also be satellite of more massive galaxies in denser regions of the simulations, and therefore not belonging to voids.

From Fig.~\ref{fig:bhmass_galmass}, BHs located in the inner regions of voids do not seem to be overmassive or undermassive with respect to the stellar mass of their galaxies. 
A selection with Voronoi cell volume leads to similar results, i.e., that BHs embedded in more or less larger Voronoi cells are not particularly under- or overmassive, but rather seem to evolve consistently with respect to their host galaxies. 
In the next section, we verify this by quantifying the BH to galaxy stellar mass ratios.

\subsubsection{BH mass to stellar mass ratios at $z=0$}
To quantify the co-evolution between the BHs and their host galaxies, we show in Fig.~\ref{fig:mass_ratio} the distribution (normalized to the total number of galaxies in the sample) of the mass ratio $M_{\rm BH}/M_{\star}$ between the BH mass and the total stellar mass of their host galaxies at redshift $z=0$. The mass ratio is an indication of whether galaxies and their central BHs co-evolve linearly, i.e., on a linear pathway in the BH mass - galaxy stellar mass diagram, or if the BHs are growing faster or slower than their host galaxies, i.e., a higher or lower mass ratio compared to the entire BH-galaxy population, respectively.
We show the mass ratio distribution for all galaxies, independently of their total stellar mass, in the left panel of Fig.~\ref{fig:mass_ratio}. Similarly, we also show the same distribution for the galaxies with $10^{9}\leqslant M_{\rm gal}< 10^{10} \, \rm M_{\odot}$ in the second panel, and finally for the most massive galaxies with $M_{\rm gal}\geqslant 10^{10}\, \rm M_{\odot}$ in the right panel. 
For all panels, the full sample of simulated galaxies is shown in grey distributions, and inner void galaxies with $r/R_{\rm void}\leqslant 0.65$ are presented as red distributions or red lines in the second panels as it only contains two void galaxies with BHs. Star symbols indicate the median of the distributions. 
We note that for the less massive galaxies of $M_{\rm gal}< 10^{9} \, \rm M_{\odot}$, 
we only find two inner void galaxies with BHs. While these BHs are under-massive compared to their hosts with respect to the average ratio distribution, we can not draw any conclusion for this mass regime based on two objects. The distributions for more massive galaxies are very similar, showing that on average, there are no under- or overmassive BHs compared to their host galaxies in voids. This suggests that mechanisms able to provide mass to galaxies also affect BH growth in the same way and that the co-evolution between void BHs and their host galaxies in voids is not strongly different from the average co-evolution in the full simulation box. We report the median and $\pm \sigma$ of the distributions in table~\ref{table:table_mass_ratio}. We also report the same quantities for slightly different void galaxy selections including both a criterion on void-centric distance and a criterion on the Voronoi volume of the galaxy cells. The mass ratios are slightly lower for $r/R_{\rm void}\leqslant 0.80$ and $V_{\rm gal}>10\, \rm (cMpc/h)^{3}$, i.e. when enforcing that galaxies are isolated, but again very similar when considering even more inner galaxies with $r/R_{\rm void}\leqslant 0.65$ (last row of Table~\ref{table:table_mass_ratio}).

\subsection{BH formation and growth history in cosmic voids with time}
\subsubsection{BH growth with time}
\begin{figure}
\centering
\includegraphics[scale=0.55]{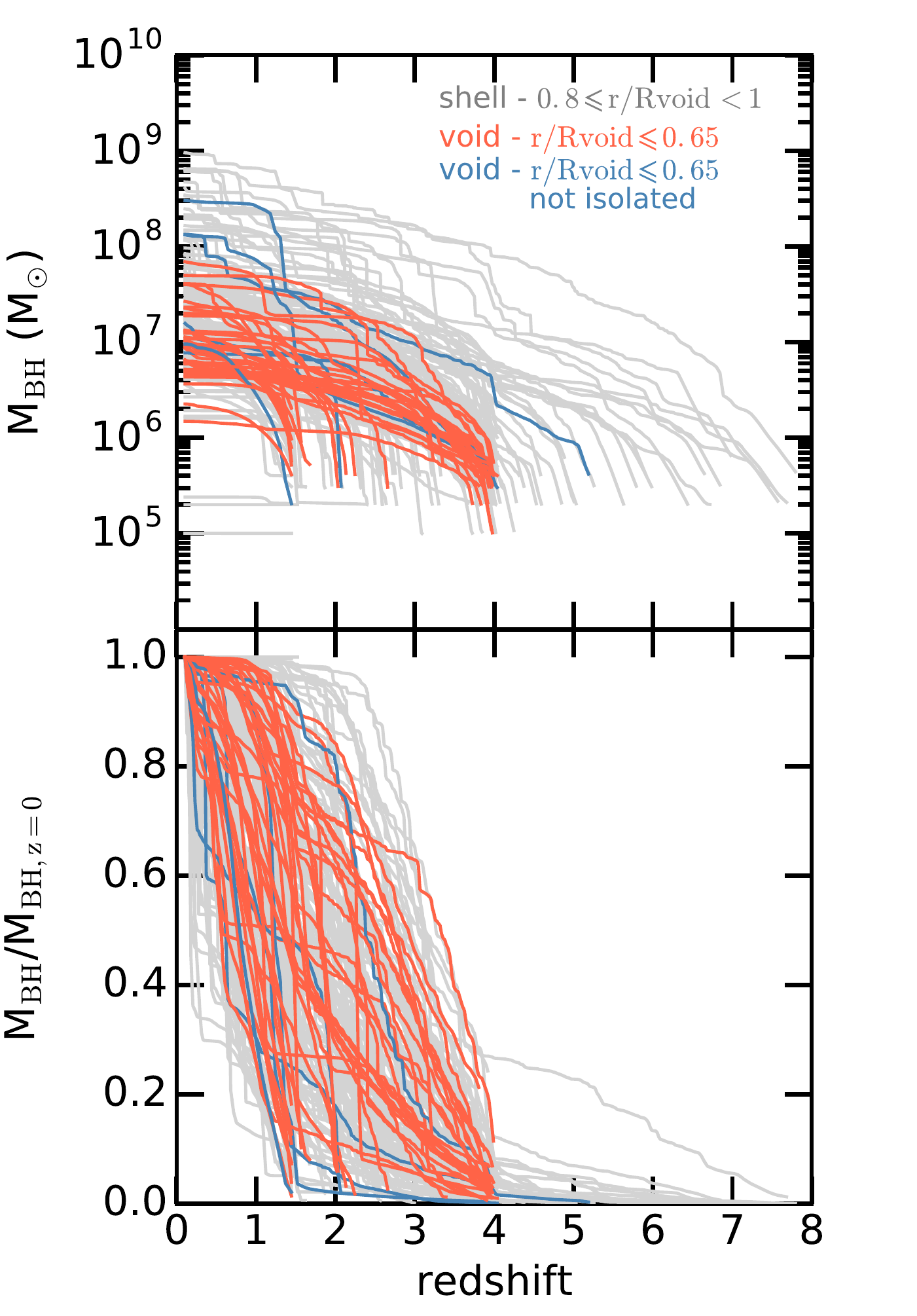}
\caption{BH growth with time, for BHs in void shell galaxies with $0.8 \leqslant r/R_{void}\leqslant 1$ (grey lines), and inner void with $\rm r/R_{void}\leqslant 0.65$ (red/blue). The bottom panel shows the BH mass growth normalized by the final mass of the BHs. Most BHs located in the inner regions of voids have a hard time to grow, similarly to their host galaxies. The three inner void galaxy BHs that are able to grow to $M_{\rm BH}\geqslant 10^{8}\, \rm M_{\odot}$ by $z=0$ (blue lines) appear to have a very small volume of their Voronoi cell ($V_{\rm gal}\leqslant 0.20\, \rm (cMpc/h)^3$), and therefore are not isolated in the cosmic web but rather seating in denser regions. Their void-centric distances show that these three BHs are not located extremely deep into their voids, which is consistent with expectations.
They illustrate the non-spherical shape of the small voids that we have identified in Horizon-AGN, and the limitation of selecting void galaxies only based on their normalized void-centric distances ($\rm r/R_{void}$).}
\label{fig:bh_growth}
\end{figure}

Do BHs in the inner regions of voids evolve more slowly with respect to the average evolution of BHs in the simulation? To understand this aspect of the BH population properties, we now focus our investigation on the growth history of BHs. We follow back in time all the central BHs of $z=0$ galaxies toward higher redshifts. 
In Fig.~\ref{fig:bh_growth}, we show the corresponding growth of BHs embedded in the shell galaxies (grey lines) and inner void galaxies with $r/R_{\rm void}\leqslant 0.65$ (red and blue lines). The bottom panel of Fig.~\ref{fig:bh_growth} shows the mass ratio of the BH mass at any given time and the final mass at $z=0$. Fig.~\ref{fig:bh_growth_vol} shows BH growth in voids with a selection based on Voronoi cell volume.

When compared to the mean evolution of BH growth in the entire simulation (not shown here), the void BHs evolve more quietly over time, and do not reach very high BH mass by $z=0$. Most of the inner void BHs do not grow efficiently. BHs located in the shells of cosmic voids (grey lines) have more efficient growth, although with a large diversity of growth histories.
The mean final mass of the inner BHs is $\langle M_{\rm BH}\rangle=2.7 \times 10^{7}\, \rm M_{\odot}$, with a median and 15th (bottom value) and 85th (top value) percentiles of $\tilde{M}_{\rm BH}=1.1 \times 10^{7}\, \rm M_{\odot}\,\, _{5.1\times10^{6}\,\rm M_{\odot}}^{4.0\times 10^{7}\,\rm M_{\odot}}$. Shell BHs end up more massive on average with a mean of $\langle M_{\rm BH}\rangle=5.7 \times 10^{7}\, \rm M_{\odot}$, and a median of $\tilde{M}_{\rm BH}=1.6 \times 10^{7}\, \rm M_{\odot}\,\, _{5.9\times 10^{6}\,\rm M_{\odot}}^{6.6\times 10^{7}\,\rm M_{\odot}}$. \\

We highlight the growth of three BHs that were able to grow to $M_{\rm BH}\geqslant 10^{8}\, \rm M_{\odot}$ with the three top blue lines in Fig.~\ref{fig:bh_growth}, and report their properties at $z=0$ in Table~\ref{table:bh_growth}. 
These three BHs appear not to be embedded in the most isolated void galaxies. Indeed, the volume of their Voronoi cell is small, with $V_{\rm gal}\leqslant 0.20 \, \rm (cMpc/h)^{3}$ for these three given BHs. 
These BHs are not isolated but, because we chose a cut $\rm r/R_{void}\leqslant 0.65$, they are labelled as inner void BHs. Their void-centric distances of, respectively 0.64,0.55 and 0.63, pass the cut only tightly. This shows on one side that, if more voids were available, even more conservative cuts than $\rm r/R_{void}\leqslant 0.65$ could be used (e.g. $\rm r/R_{void}\leqslant 0.55$)  to increase reliability. On the other side, it also shows the limitations of relying only on the $r/R_{void}$ cut of the void-centric distance.
These distances have been computed assuming that voids are spherical entities, and the present example illustrates the non-spherical shape of voids, and its impact when selecting void galaxies  within a given void-centric distance only. These BHs that are located close to or within denser regions of the cosmic web have experienced rapid and major growth events, for two of them, that can be related to galaxy mergers and central BH-BH mergers. The last BH has formed quite early ($z\sim 5.5$), which is not the case for the vast majority of the void BHs in our study, as explained below.
This work highlights that selecting voids by considering conservative and multiple criteria is optimal to ensure a robust selection, especially in the case of low statistics. Stacking voids in large samples of cosmic voids is a powerful way to enhance robustness: it allows for a more conservative choice of what we define as ``inner'' void galaxies, and reduces the impact of the effects discussed above.

\begin{table}
\caption{Properties of the inner void BHs with $M_{\rm BH}\geqslant 10^{8}\, \rm M_{\odot}$ at $z=0$. The BH host galaxies have a very low volume of their Voronoi cells, indicating that they are not particularly isolated in the cosmic web but rather seating in denser regions. Based on the void-centric distances, we also note that these galaxies are not located extremely deep into their voids, which is consistent with expectations.}
\begin{center}
\begin{tabular}{ccc}
\hline
$M_{\rm BH}$ (M$_{\odot}$) & Volume Voronoi $\rm (cMpc/h)^{3}$ & $r/R_{\rm void}$\\
\hline
$3.04\times 10^{8}$ & $0.18$ & $0.64$\\
$1.34\times 10^{8}$ & $0.19$ & $0.55$\\
$1.33\times 10^{8}$ & $0.04$ & $0.63$\\
\hline
\end{tabular}
\end{center}
\label{table:bh_growth}
\end{table}

\subsubsection{BH formation time}
\begin{table*}
\caption{Mean formation time $\langle t_{\rm form}\rangle$ of the inner void BHs ($r/R_{\rm void}\leqslant 0.65$, first row), void BHs ($r/R_{\rm void}<0.80$, second row), and BHs located in void shell galaxies ($r/R_{\rm void}$, third row), for different stellar mass bins. We also show the percentage of BHs that first appear in the simulation before redshift $z=4$ ($t_{z>4}$). The choice of this redshift is arbitrary, and only used to provide some insights when we compare to the full simulation (last row). 
BHs embedded in more massive galaxies tend to form earlier in both the voids and in the full simulation. BHs that are located deeper into the voids have formed more recently on average, compare to BHs located in the shells (which represent denser regions). For the same galaxy stellar mass bins, the BHs in the inner regions of voids have, on average, formed more recently than the global population of BHs in the simulation.}
\begin{center}
\begin{tabular}{lcccc}
\hline
 & $M_{\rm gal}=10^{8-9}\, \rm M_{\odot}$ & $10^{9-10} \, \rm M_{\odot}$ & $10^{10-11}\, \rm M_{\odot}$ & $10^{11-12}\, \rm M_{\odot}$\\
\hline
\hline
mean  $\langle t_{\rm form} \rangle$ in inner gal.  & $2.2$ & $3.0$ & $2.8$ & $4.3$ \\
percentage inner gal. with $t_{z>4}$ & $0\%$ & $0\%$ & $0\%$ & $50\%$\\
\hline
mean  $\langle t_{\rm form} \rangle$ in all void gal. & $2.5$ & $2.7$ & $3.1$ & $4.9$\\
percentage void gal. with $t_{z>4}$ & $0\%$ & $0\%$ & $9.7\%$ & $63\%$\\
\hline
mean  $\langle t_{\rm form} \rangle$ in shell gal.
& $2.3$ & $2.5$ & $3.5$ & $4.7$  \\
percentage shell gal. with $t_{z>4}$ & $0\%$ & $1.4\%$ & $9.80\%$ & $58\%$ \\
\hline
full simulation $t_{z>4}$ & $1.8\%$ & $7.0\%$ & $35.3\%$ & $82.7\%$ \\
\hline
\end{tabular}
\end{center}
\label{table:bh_formationtime}
\end{table*}

In this section, we investigate the initial assembly of the BH population. To do so, we trace back in time the BH sink particles to higher redshifts until we reach the time of the first appearance in the simulation. In the following, we name this time {\it the BH formation time} ($t_{\rm form}$). This time is useful to get a sense of whether the void BHs form at early times or, instead, at pretty late times compared to the BHs in denser regions.
As before, we divide the population of void BHs in three samples: inner void BHs ($r/R_{\rm void}\leqslant 0.65$), void BHs ($r/R_{\rm void}\leqslant 0.80$), and BHs located in shells ($0.80\leqslant r/R_{\rm void}\leqslant 1$). 

We provide the mean formation time $t_{\rm form}$ of these samples in Table~\ref{table:bh_formationtime} for different bins of the BH host galaxy stellar mass, i.e. for $M_{\rm gal}=10^{8-9}\, \rm M_{\odot}$, $10^{9-10}\, \rm M_{\odot}$, $10^{10-11}\, \rm M_{\odot}$, $10^{11-12}\, \rm M_{\odot}$. 
On average, BHs embedded in more massive galaxies have formed at earlier times; this is true for both the entire simulation and the voids.
To understand whether inner void BHs generally form earlier/later in voids than in other regions of the simulation, we look at the fraction of BHs that has been formed before a given threshold in time for both the void BHs and the full simulation. We choose this threshold to be redshift $z=4$ \footnote{We note here that at this time, a new level of refinement of the simulation grid is created in Horizon-AGN. Gas cells are more resolved, which can boost the formation of stellar and BH particles.}. We show the fraction of BHs that have been formed before $z=4$ in Table~\ref{table:bh_formationtime}, for all our sample of BHs.
We find that BHs embedded in voids form on average at later times than for the full simulation. The effect is enhanced for BHs located in inner void regions. 
For example, $35\%$ of the galaxies in the stellar mass bin $M_{\rm gal}=10^{10-11}\, \rm M_{\odot}$ of the full simulation host BHs that have formed earlier than $z=4$. However, only $\sim 10\%$ of the void BHs were present at that time, and none of the inner void BHs.

In Horizon-AGN, BHs of $10^{5}\, \rm M_{\odot}$ are placed in gas cells with a density higher than the density threshold to form stars (i.e., $\rho>\rho_{0}$) if no other BH has been formed within 50 kpc. The seeding of the simulation is therefore somehow related to the assembly of the galaxies. 
The apparent late assembly of the inner void BHs could be intertwined to the assembly of their host galaxies.

\subsection{BH accretion rate: do void galaxies host AGN?}
In Section 4, we have seen that the fraction of star-forming galaxies among inner void galaxies is higher than the average fraction of the entire simulated volume for $r/R_{\rm void}\leqslant 0.7$ (Fig.~\ref{fig:fraction_sf}). Dense cold gas available to form new stars could also fuel existing central BHs. Early works have suggested that the growth of BHs should be closely linked to the assembly of the host galaxies \citep{Silk1998,Kauffmann2000}. Correlations between star formation activity and accretion onto BHs are still unclear and subject to debate today \citep[e.g.,][]{2012MNRAS.419...95M,2013ApJ...773....3C,2015ApJ...800L..10R,2015ApJ...806..187A,2019MNRAS.484.4360A}.

Dense environments with galaxy-galaxy interaction and mergers could also favor the gas supply to the center of galaxies and trigger new star formation episodes and fuel the central BHs \citep[][and references therein]{2017MNRAS.469.4437C,2019MNRAS.484.4413H,2018MNRAS.479.4056W}. 
However, here we aim at understanding these correlations in under-dense environments where the accretion onto the BHs should not be driven substantially by galaxy-galaxy mergers and interactions with the surroundings. 
Observational studies have shown that the fraction of strong-lined AGN could be as twice as numerous (at fixed galaxy mass) in under-dense regions with respect to denser regions \citep{Kauffmann2003}, although with a nonconservative definition of under-dense regions.  
More recently and based on SDSS DR2 analysis, \citet{2008ApJ...673..715C} show that AGN are common in cosmic voids. Their study investigates how the accretion properties of different types of spectrally defined AGN (LINERS, Seyferts, Transition objects) correlate with other BH properties and properties of their host galaxies.

\begin{figure*}
\centering
\includegraphics[scale=0.43]{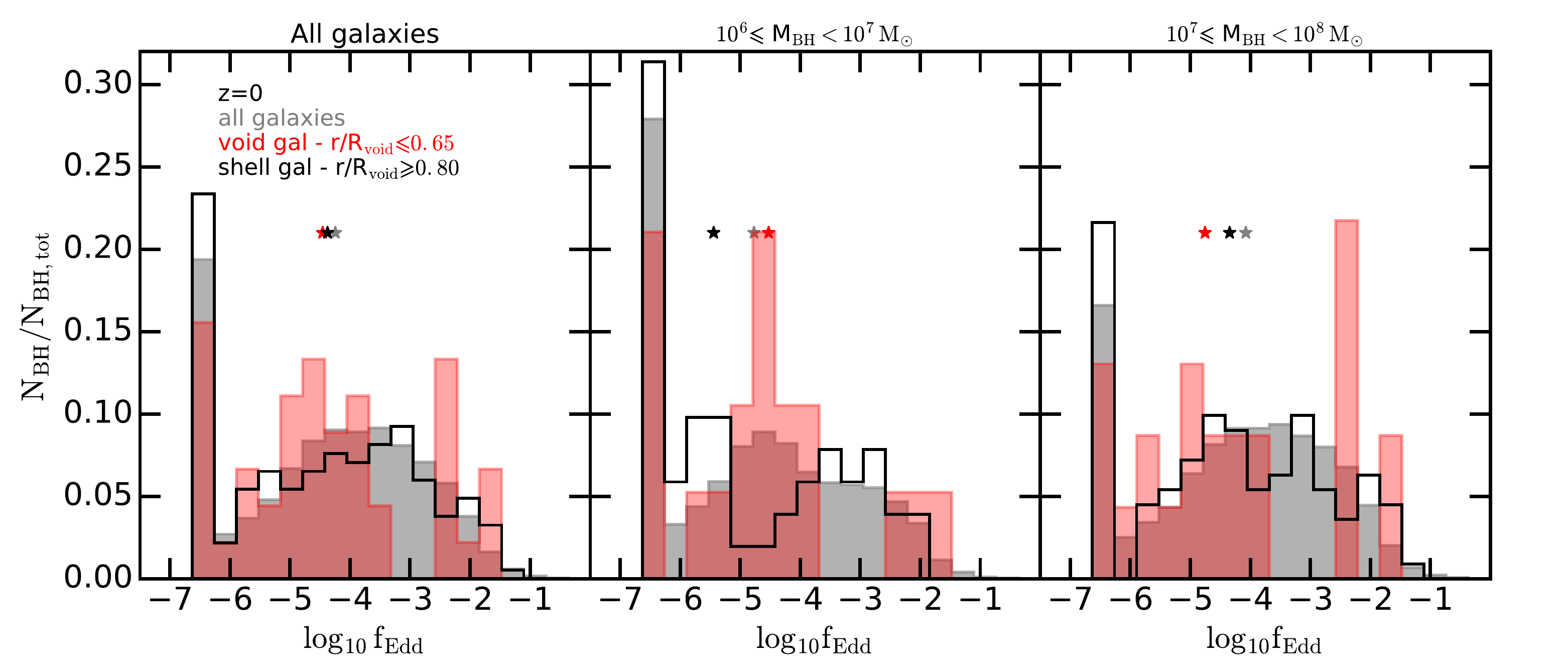}\\
\includegraphics[scale=0.43]{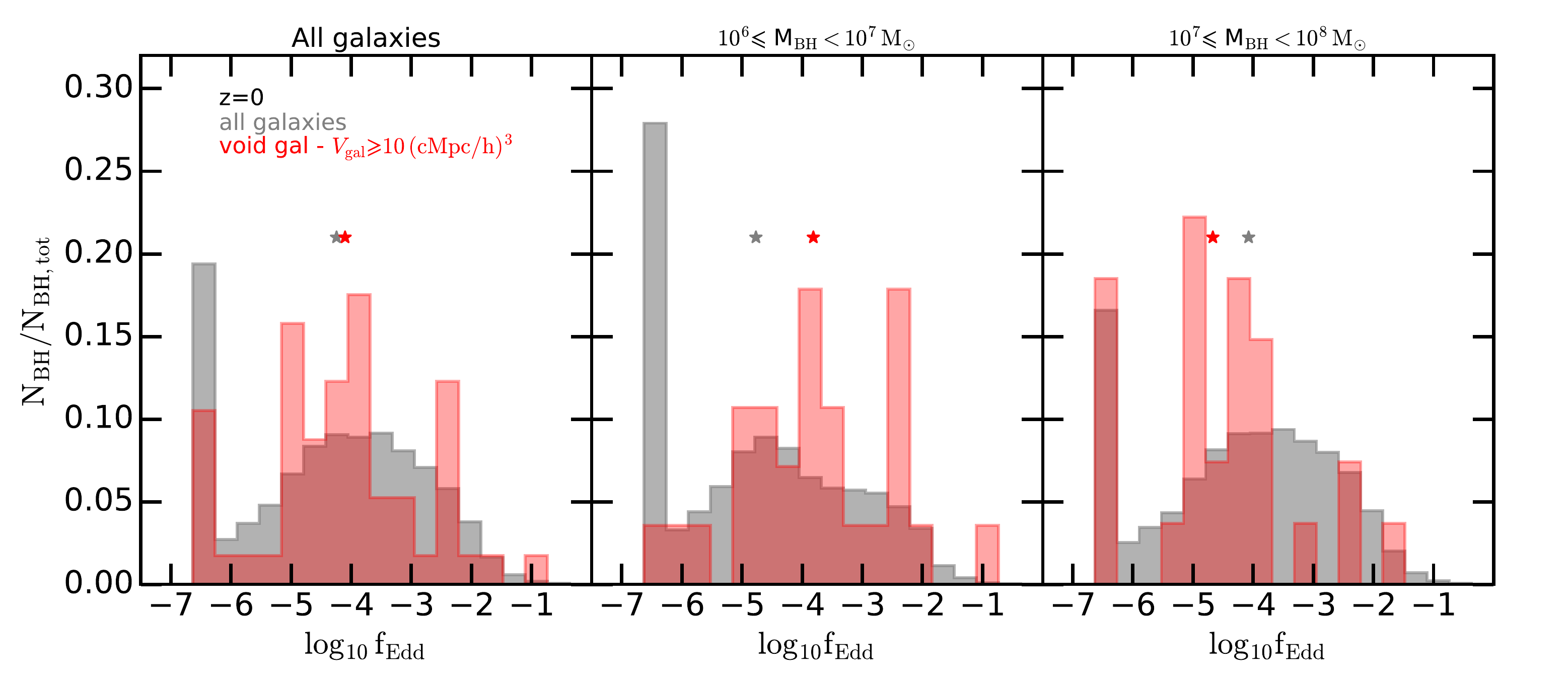}\\
\caption{Eddington ratio distributions for different BH mass bins (left panels: all BH mass, middle panels:$10^{6}\leqslant M_{\rm BH}< 10^{7}\, \rm M_{\odot}$, left panels: $10^{7}\leqslant M_{\rm BH}< 10^{11}\, \rm M_{\odot}$). Top panels include a selection with void-centric distances (red: $r/R_{\rm void}\geqslant 0.65$, black: $0.8\leqslant r/R_{\rm void}\leqslant 1$), whereas bottom panels use the volume of the Voronoi cells around void galaxies ($V_{\rm gal}\geqslant 10\, \rm (cMpc/h)^{3}$). For visual purposes, we set $\log_{10}\, f_{\rm Edd}=-6.5$ for all BHs with lower Eddington ratios. Star symbols indicate the median of the distributions. }
\label{fig:bh_activity}
\end{figure*}

\subsubsection{BH Eddington ratios}
To study the ability of BHs to accrete the surrounding gas, we often use the Eddington ratio quantity, which is defined as the ratio between the accretion rate onto the BHs and their Eddington accretion rate: $f_{\rm Edd}=\dot{M}_{\rm BH}/\dot{M}_{\rm Edd}$. 
We can consider a BH as being efficiently accretion BH, the so-called AGN, if $\log_{10}\dot{M}_{\rm BH}/\dot{M}_{\rm Edd}\geqslant -3$.
In Fig.~\ref{fig:bh_activity}, we show the distribution of BH Eddington ratios at $z=0$ in all galaxies (left panels), galaxies hosting BHs of $10^{6}\leqslant M_{\rm BH}<10^{7}\, \rm M_{\odot}$ (middle panels) and BHs of $10^{7}\leqslant M_{\rm BH}< 10^{11}\, \rm M_{\odot}$ (right panels). Grey histograms represent the distribution of all simulated galaxies (located in voids or not), black histograms the distribution of Eddington ratios in shell galaxies (defined as $r/R_{\rm void}>0.80$), and red histograms show either the distribution of central void galaxies with $r/R_{\rm void}\leqslant 0.65$ (top panels), or isolated galaxies with large Voronoi cell volume of $V_{\rm gal}\geqslant 10 \rm (cMpc/h)^{3}$ (bottom panels).

However, we find bimodal distributions for the inner void galaxies with or without a selection in BH mass. While a large fraction of BHs have low Eddington ratios with $\log_{10} f_{\rm Edd}\leqslant -3$, there is a given fraction of AGN with $\log_{10} f_{\rm Edd}\geqslant -3$ in all panels. From a KS test, we can not rule out that the distribution of Eddington ratios for the inner void BHs is not drawn from the distribution of the entire simulation (KS test: $p\geqslant 0.3$). These results need to be carefully checked with larger samples, as the limited number of inner void BHs is below the sanity number of objects for a KS test. 
At $z=0$, the distribution for all galaxies (in grey) peaks around $\log_{10} f_{\rm Edd}\sim -4$. The distributions of the shell galaxies follow the distributions of all galaxies, independently of BH mass selection (as expected since the shells constitute the walls of voids, and therefore are denser environments). 
We show the same distribution but for isolated galaxies with large Voronoi cell volume (red) in the bottom plots of Fig.~\ref{fig:bh_activity}. These distributions, which include a larger number of objects, are statistically different (KS test: $p\leqslant0.04$) from the distributions of the shells or of the entire simulation for $10^{6}\leqslant M_{\rm BH}\leqslant 10^{7}\, \rm M_{\odot}$ and $10^{7}\leqslant M_{\rm BH}\leqslant 10^{8}\, \rm M_{\odot}$. 
In any case, our work demonstrates that at least some of the inner void regions or the isolated regions of the cosmic web host accreting AGN.

\begin{table*}
\caption{Percentage of AGN with $\log_{10} f_{\rm Edd}\geqslant -3$ in the inner region of cosmic void (red distributions in Fig.~\ref{fig:bh_activity}, $r/R_{\rm void}\leqslant 0.65$), voids ($r/R_{\rm void}\leqslant 0.8$), shell galaxies (black distributions, $0.8\leqslant r/R_{\rm void} \leqslant 1$), and in the entire simulation for comparison (grey distributions). The size of the samples is given in parentheses.
We compute the percentages for several samples binned by BH mass: all BHs, BHs with $10^{6}\leqslant M_{\rm BH}\leqslant 10^{7}\, \rm M_{\odot}$, and BHs with $10^{7}\leqslant M_{\rm BH}\leqslant 10^{8}\, \rm M_{\odot}$. We also show the AGN fraction for different samples of galaxies selected on their Voronoi cell volume: $V_{\rm gal}\geqslant 1 \rm (cMpc/h)^{3}$ (the distribution is similar to the grey distribution below, in Fig.~\ref{fig:bh_activity}), $V_{\rm gal}\geqslant 10\, \rm (cMpc/h)^{3}$ and $V_{\rm gal}\geqslant 20 \,\rm (cMpc/h)^{3}$. We find similar or higher fractions of AGN in voids or their inner regions than in the entire simulation. Similar results are found for isolated galaxies ($V_{\rm gal}\geqslant 10\, \rm (cMpc/h)^{3}$), at least for two of the BH mass ranges.}
\begin{center}
\begin{tabular}{lllll}
\hline
& all BHs & $M_{\rm BH}=10^{6-7}\, \rm M_{\odot}$ & $M_{\rm BH}=10^{7-8}\, \rm M_{\odot}$ & $M_{\rm BH}= 10^{8-9}\, \rm M_{\odot}$\\
\hline
all galaxies & $20 \%$ & $16 \%$ & $23 \%$ & $16\%$\\
\hline
inner void BHs & $22 \% \,(10/45)$ & $16 \% \,(3/19)$ & $30 \% \,(7/13)$ & $0\% \,(0/3)$\\
void BHs & $24 \% \,(32/136)$ & $18 \% \,(8/45)$ & $29 \% \,(22/176)$ & $15\% \,(2/13)$\\
shell BHs & $19 \%\, (35/184)$ & $16 \%\,(8/51)$ & $22 \%\,(24/111)$ & $14\%\,(3/21)$\\
\hline
$V_{\rm gal}\geqslant 1\, \rm (cMpc/h)^{3}$ & $22\%\, (77/346)$ & $20\% \,(24/119)$ & $24\%\, (51/209)$ & $11\% \,(2/18)$\\
$V_{\rm gal}\geqslant 10\, \rm (cMpc/h)^{3}$ & $21\%\, (12/57)$ & $29\% \,(8/28)$ & $11\%\, (3/27)$ & $50\% \,(1/2)$\\
$V_{\rm gal}\geqslant 20\, \rm  (cMpc/h)^{3}$ & $11\% \,(1/9)$ & $0\%\, (0/5)$ & $25\% \,(1/4)$ & $0\%\, (0/0)$\\
\hline
\end{tabular}
\end{center}
\label{table:bh_agn_fraction}
\end{table*}

\begin{table*}
\caption{Mean galaxy stellar mass (in $\rm M_{\odot}$) of the void AGN host galaxies located in the inner region of voids ($r/R_{\rm void}\leqslant 0.65$), in voids ($r/R_{\rm void}\leqslant 0.80$), in the shells ($0.80\leqslant r/R_{\rm void}\leqslant 1$). For comparison, we also provide the mean stellar mass of the galaxies hosting AGN in the entire simulation (first row).}
\begin{center}
\begin{tabular}{ccccc}
\hline
& all BHs & $M_{\rm BH}=10^{6-7}\, \rm M_{\odot}$ & $M_{\rm BH}=10^{7-8}\, \rm M_{\odot}$ & $M_{\rm BH}= 10^{8-9}\, \rm M_{\odot}$\\
\hline
all galaxies & $3.2\times 10^{9}$ & $6.1\times 10^{8}$ & $2.5\times 10^{9}$ & $1.2 \times 10^{10}$\\
\hline
$r/R_{\rm void}\leqslant 0.65$ & $1.0 \times 10^{9}$ & $4.8 \times 10^{8}$ & $1.3 \times 10^{9}$ & - \\
$r/R_{\rm void}\leqslant 0.80$ & $2.1 \times 10^{9}$ & $5.5 \times 10^{8}$ & $1.9 \times 10^{9}$ & $1.0 \times 10^{10}$ \\
shell BHs                      & $3.0 \times 10^{9}$ & $5.9 \times 10^{8}$ & $2.6 \times 10^{9}$ & $1.2 \times 10^{10}$ \\
\hline
\end{tabular}
\end{center}
\label{table:agn_host}
\end{table*}

We summarize in Table~\ref{table:bh_agn_fraction} the fraction of AGN in all simulated galaxies and void galaxies.
There is an increasing fraction of AGN in more and more massive galaxies (i.e., from the column $M_{\rm BH}=10^{6}-10^{7}\, \rm M_{\odot}$ to $M_{\rm BH}=10^{7}-10^{8}\, \rm M_{\odot}$) up to the corresponding galaxy mass $M_{\rm gal}\leqslant 3\times 10^{10}\, \rm M_{\odot}$. 
This is true for the entire simulation, but also the different void selections.
For example, the fraction of AGN among the inner void galaxies jumps from $16\%$ for $M_{\rm BH}=10^{6-7}\, \rm M_{\odot}$ to $30\%$ for the $M_{\rm BH}=10^{7-8}\, \rm M_{\odot}$ BHs.
We find lower fractions among the most massive galaxies and correspondingly BHs with $M_{\rm BH}= 10^{8-9}\, \rm M_{\odot}$, in agreement with the trend for the entire simulated volume. This is a regime that would be affected by quenching, i.e., the suppression of star formation sustained in time, caused by AGN feedback. In cosmological simulations, AGN feedback also often ceases the accretion onto the BHs themselves \citep{2019MNRAS.484.4413H}. We find no AGN in the inner void galaxies of this mass range. Massive galaxies with these masses are rare in our void sample, and therefore our findings may suffer strongly from the very low number statistics in this mass regime.

More importantly, we find that there are more AGN among void BHs ($r/R_{\rm void}\leqslant 0.80$) and inner void BHs ($r/R_{\rm void}\leqslant 0.65$) than in the shells. 
For example, we find that $29\%$ of void BHs with $M_{\rm BH}=10^{7-8}\, \rm M_{\odot}$ are AGN, compared to $22\%$ for the shell BHs, and $23\%$ for same mass BHs of the entire simulation.

We now turn to look in more detail at those AGN in voids to understand what are their properties and where they are located inside the voids.
We find that the inner void AGN have a mean void-centric distance of $\langle r/R_{\rm void}\rangle=0.5$ and a mean volume of the Voronoi cell of $\langle V_{\rm gal}\rangle=10 \, \rm (cMpc/h)^{3}$ for $M_{\rm BH}=10^{6-7}\, \rm M_{\odot}$, and $\langle r/R_{\rm void}\rangle =0.6$ and $\langle V_{\rm gal}\rangle =16 \, \rm (cMpc/h)^{3}$ for more massive BHs with $M_{\rm BH}=10^{6-7}\, \rm M_{\odot}$. Therefore these AGN tend to be located in the inner region of cosmic voids, but also in isolated galaxies.

Similarly, we find that isolated galaxies (arbitrarily defined as $V_{\rm gal}\geqslant 10 \, \rm (cMpc/h)^{3}$ given the distribution of the volume of the Voronoi cells) in voids definitely have a high fraction of AGN (with $\log_{10} f_{\rm Edd}\geqslant -3$), particularly for 
$10^{6}\leqslant M_{\rm BH}\leqslant 10^{7} \, \rm M_{\odot}$ with $29\%$ of AGN. 
For our selection on Voronoi cell volume we do not require the galaxies to be located within $r/R_{\rm void}$. In fact, we find that the mean void-centric distance of the AGN in isolated galaxies is $\langle r/R_{\rm void}\rangle=0.92$ for BHs in the mass range $M_{\rm BH}=10^{6-7}\, \rm M_{\odot}$, and $\langle r/R_{\rm void}\rangle =0.78$ for more massive BHs in the range $M_{\rm BH}=10^{7-8}\, \rm M_{\odot}$. 
Therefore, some of the AGN are located quite far from the spherically defined center of the voids while still being in isolated galaxies. 
Again, our results highlight here the risk of using only normalized void-centric distances in stacked observational samples, where void galaxies could be misjudged as not being part of voids. 
This reaffirms the power of using both void-centric distances and Voronoi cell volume information to build samples of void galaxies---both methods promise to be even more powerful with a larger sample of voids.

To conclude about AGN in voids, a third of the BHs in inner regions of voids could be detected as AGN with $\log_{10} f_{\rm Edd}\geqslant -3$. Therefore, at least some of the void galaxies that are on the star-forming main sequence are also able to fuel their central BHs with Eddington ratios of $\log_{10} f_{\rm Edd}\geqslant -3$.
To provide some ideas of the luminosity of these AGN in cosmic voids, we have computed their hard X-ray luminosities (2-10 keV) using the bolometric correction of \citet{Hopkins2007}.
We find that the inner void AGN of $M_{\rm BH}=10^{7}-10^{8}\, \rm M_{\odot}$ with $\log_{10} f_{\rm Edd}\geqslant -3$ have a mean X-ray luminosity of $\langle L_{2-10} \rangle=41.9 \, \rm erg/s$, the less luminous AGN being $L_{2-10}=41.5\, \rm erg/s$ and the most luminous $L_{2-10}=42.4\, \rm erg/s$.
For less massive inner void AGN with $M_{\rm BH}=10^{6}-10^{7}\, \rm M_{\odot}$ we find  $\langle L_{2-10} \rangle=41.7 \, \rm erg/s$ and $L_{2-10}=41.5, 42.0\, \rm erg/s$ for the less and most luminous AGN, respectively.

\subsubsection{Galaxy hosts of the void AGN}
We provide in Table~\ref{table:agn_host} the mean galaxy mass of the AGN that we discussed above, for the different BH mass bins.
We find that the host galaxies of the void AGN have, on average, masses very similar or slightly lower than the AGN galaxy hosts in the entire simulated volume. 
The AGN host galaxies that we identify in the shells are even more consistent with the AGN hosts in the total simulated volume, as expected since the shells of voids represent denser regions of the cosmic web.
For example, the AGN with BH mass in the range $M_{\rm BH}=10^{6-7}\, \rm M_{\odot}$, have a mean galaxy stellar mass of $\langle M_{\rm gal}\rangle = 4.8 \times 10^{8}\, \rm M_{\odot}$ if located in the inner void regions, $\langle M_{\rm gal}\rangle = 5.9 \times 10^{8}\, \rm M_{\odot}$ in the shells. For comparison, we find $\langle M_{\rm gal}\rangle = 6.1 \times 10^{8}\, \rm M_{\odot}$ on average for the simulation. 
We conclude that the AGN that we observe in the void shells are not hosted by particularly different galaxies than in the entire simulation. However, the host galaxies of inner void AGN have lower masses than the average of the simulation.

Galaxies with signatures of AGN activity have been found in observations in moderately bright and moderately massive galaxies in SDSS voids, i.e., with $M_{\rm r}<20$ and $M \sim 10^{10.5}\, \rm M_{\odot}$ \citep{2008ApJ...673..715C}. However, the fraction of AGN in the most massive ($M > 10^{10.5}\, \rm M_{\odot}$) or bright galaxies ($M_{\rm r}\sim 20.5$) was found to be lower. This has been confirmed recently in \citet{2018A&A...612A..31P}, where they use a 3D high resolution (2.6Mpc/h) density field obtained from a Bayesian reconstruction applied to the M++ galaxy catalog to identify different environments, coupled to the MPA/JHU AGN catalogs \citep{Kauffmann2003,Brinchmann+04}. Larger simulations (larger voids, more statistics) will be needed to capture the evolution of the accretion properties in void galaxies more massive than what we can do with Horizon-AGN.


\section{Conclusions}
The analyses of observational samples of void galaxies have shown that these galaxies present different properties in terms of colors, morphologies, sizes of stellar radii, metallicity, compared to galaxies embedded in denser environments of the cosmic web \citep{2004ApJ...617...50R,2005ApJ...624..571R,2006MNRAS.372.1710P,2008MNRAS.384.1189V,2012MNRAS.426.3041H,2012AJ....144...16K,2014MNRAS.439.3564C}.
Their evolution could also be ``late'' \citep{2014MNRAS.445.4045R}.
In this paper, we have used the state-of-the-art cosmological hydrodynamical simulation Horizon-AGN (100 cMpc side length) to investigate for the first time and statistically the properties of galaxies and BHs in cosmic voids. We have used the \texttt{VIDE} toolkit \citep{2015A&C.....9....1S}, a public code based on \texttt{ZOBOV} \citep{2008MNRAS.386.2101N}, to identify the cosmic voids. The code uses tracers, in our case galaxies, to reconstruct the density field and select under-dense regions. Therefore, the voids are not defined as empty regions of the cosmic web, but rather as under-dense regions that contain a few void galaxies. Our method mimics what is achievable in observations.
We summarize our main results below. 

\begin{itemize}
\item We have compared two approaches to study the properties of galaxies and BHs in cosmic voids: void-centric distances and the degree of isolation of the galaxies (i.e., with the volume of their Voronoi cell).
We find that using only the void-centric distances can lead to the misidentification of inner void galaxies in small and non-spherical voids. 
Our work highlights the importance of conservative cuts and of the quality of void definition, enhanced by combining void-centric distance and Voronoi cell volume to select galaxies embedded in very low density regions accurately. This is particularly relevant when studying small cosmic voids and when the statistics is low.

\item We find a higher relative abundance of low-mass galaxies in the center of cosmic voids (Fig.~\ref{fig:distri_gal_voidcentric}). Interestingly, at fixed halo mass and for halos of $M_{\rm h}\leqslant 10^{10.75} \,\rm M_{\odot}$, we find that the mean stellar mass of the inner void galaxies is lower than the mean for the entire simulation (Fig.~\ref{fig:mstar_mhalo}).
We find similar results for the isolated galaxies with Voronoi cell volume of $V_{\rm gal}\geqslant 10\, \rm (cMpc/h)^3$ and $M_{\rm h}\leqslant 10^{11.25} \,\rm M_{\odot}$. Inner void and isolated galaxies have a harder time growing in stellar mass compared to other galaxies embedded in the same-mass halos in denser regions of the cosmic web. 

\item Regarding the star-forming properties of void galaxies, we find that they have, on average, lower SFR (Fig.~\ref{fig:distri_sfr_rad}), but slightly higher specific star formation rate sSFR (Fig.~\ref{fig:distri_sfr_rad}). Therefore, at least some of the void galaxies form stars more efficiently with respect to their mass. Moreover, the fraction of star-forming galaxies is higher in void centers with respect to the fraction in the entire simulation (Fig.~\ref{fig:fraction_sf}). Therefore, the small voids of Horizon-AGN seem to be relatively active regions of the cosmic web.  

\item We find that these results can not be only attributed to the relative abundance of low-mass galaxies in voids (Fig.~\ref{fig:fraction_sf}). Indeed, we obtain the same enhancement of star-forming galaxy fraction in voids for different galaxy mass bins ($M_{\rm gal}\leqslant 10^{9.5} \, \rm M_{\odot}$, and $M_{\rm gal}\geqslant 10^{9.5} \, \rm M_{\odot}$). 

\item We find that the lower SFR in void galaxies stands for all galaxy mass bins (Fig.~\ref{fig:biased_gal}). The discrepancy is very small but may indicate that the star formation activity could depend slightly on the environment. We argue that our sample of voids galaxies is too small to draw strong and definitive conclusions, but our work shows the importance of investigating this effect on larger volume simulations.

\item BHs in central regions of voids have, on average, lower masses than in denser regions of the simulation (Fig.~\ref{fig:distri_bh}). This is mainly due to the prevalence of low-mass galaxies in voids. In fact, we find that the distribution of the mass ratio between BH mass and their host galaxy stellar mass is not different from the distribution of the full simulation (Fig.~\ref{fig:bhmass_galmass}). This suggests that even if the physical processes responsible for growing galaxies and BHs that are located in cosmic voids are different than those in dense regions (number of mergers, dynamical interactions, etc), they impact the growth of the void galaxies and their central BHs in the same way.

\item We find that void BHs have, on average, formed more recently than the BHs embedded in similar mass galaxies in the entire simulation. Only the most massive BHs located in voids have formed before redshift $z>4$, while an important number of BHs in lower-mass galaxies in the entire simulation formed earlier. Our results could suggest that at least some of the galaxies and their supermassive BHs in our void sample may have assembled at later times.

\item The mass assembly of an important number of inner void BHs is slow (Fig.~\ref{fig:bh_growth}). These BHs formed at $z\sim 4$ or more recently, and only increased their mass of less than one order of magnitude by $z=0$. However, some of the BHs located in the inner region of voids ($r/R_{\rm void}\leqslant 0.65$) experienced a quite rapid growth. 
By analyzing their Voronoi cell volumes, we were able to show that these BHs are not close to the deep center of voids, they actually sit near the limit of $r/R_{\rm void}\leqslant 0.65$, and therefore are, consistently with expectations, more subject to galaxy mergers and other external processes. A more stringent selection on a larger sample of voids will allow a more robust selection.

\item We find a bimodal distribution of the BH Eddington ratios (Fig.~\ref{fig:bh_activity}), which does not exist for the distribution of all the simulated galaxies, or the distribution of shell galaxies. This bimodal distribution is also present when we divide the void BH sample into low-mass BHs with $10^{6}\leqslant M_{\rm BH} < 10^{7} \, \rm M_{\odot}$ and more massive BHs $M_{\rm BH}\geqslant 10^{7}\, \rm M_{\odot}$. We find that $22-24\%$ of void BHs could be detected as AGN with $\rm \log_{10} f_{Edd}>-3$. The fraction of AGN varies as a function of void-centric distances and BH mass.
In terms of hard X-ray (2-10 keV) luminosities, we find that the inner void galaxies with $\rm \log_{10} f_{Edd}>-3$ and $10^{6}\leqslant M_{\rm BH} < 10^{8} \, \rm M_{\odot}$ have $L_{\rm 2-10}\geqslant 10^{41.5}\, \rm erg/s$, half of them having $L_{\rm 2-10}\sim 10^{42}\, \rm erg/s$.

\end{itemize}

\section{Discussion}
Although our simulation is the largest state-of-the-art hydrodynamical simulation used so far to identify cosmic voids with a Voronoi tessellation algorithm, we are still limited by the simulated volume of Horizon-AGN and can not identify voids as large as in observations (e.g., from the SDSS survey). The mean radius of our voids is $\langle R_{\rm void} \rangle\sim 4\, {\rm cMpc}/h$, while the radii of the observed voids are of a few tens of ${\rm cMpc}/h$ \citep{2002ApJ...566..641H,2002MNRAS.330..399P,2014MNRAS.445.4045R}. Our sample of void galaxies is also relatively small to draw definite conclusions, but our results on galaxy properties are so far in good agreement with observations from SDSS. Our analysis is a first attempt to compare the results from observational surveys to a full hydrodynamical simulation, and would gain from comparing with larger simulations (e.g., the IllustrisTNG simulation of 300 cMpc side length).

\subsection{Comparisons with a current observational sample of void galaxies}
Our analysis on the properties of void galaxies, in its spirit, is very close to the study that was conducted in \citet{2014MNRAS.445.4045R} with the SDSS DR7 data.
In this section, we compare our results from the cosmological simulation Horizon-AGN to their results. 

In the observational sample used in \citet{2014MNRAS.445.4045R}, low-mass galaxies are also more abundant in voids rather than within their control field, or shell regions defined as galaxies out to $\leqslant 30\, h^{-1} \rm  Mpc$, which corresponds to  $r/R_{\rm void}\geqslant 1.5$. Even if our limits to define shell galaxies do not use the exact same values of $r/R_{\rm void}$, they actually both correspond to the void-centric radius at which we observe the highest number of galaxies, namely what is commonly referred to as the wall of the void ($r/R_{\rm void}=1$ for our catalog, and $r/R_{\rm void}=1.5$ in \citet{2014MNRAS.445.4045R}).
We find a very good agreement for the star-forming properties of the simulated void galaxies (Fig.~\ref{fig:distri_sfr_rad}, top panel) and in the observations.
The observed SDSS void galaxies have lower SFR when approaching the center of voids, from $r/R_{\rm void}=1.5$ to $r/R_{\rm void}=0.3$, with a slight increase for $r/R_{\rm void}=0.2$ but still with lower SFR than denser regions \citep{2014MNRAS.445.4045R}. Similarly, they find slightly higher specific star formation rates for lower void-centric radii, as we do (bottom panel of Fig.~\ref{fig:distri_sfr_rad}). It indicates that, on average, the void galaxies could be able to form stars slightly more efficiently with respect to their mass.

Some discrepancies arise from the star-forming fraction as a function of void-centric radius for different stellar mass bins. In Fig.~\ref{fig:fraction_sf}, we have shown that the fraction of star-forming void galaxies was higher for more massive galaxies of $M_{\rm gal}\geqslant 10^{9.5}\, \rm M_{\odot}$, than for galaxies of $M_{\rm gal}< 10^{9.5}\, \rm M_{\odot}$.
However, \citet{2014MNRAS.445.4045R} find the opposite trend in the SDSS catalog, i.e., that galaxies of $M_{\rm gal}\geqslant 10^{9.5}\, \rm M_{\odot}$ have lower SFR than the less massive galaxies.
When only selecting galaxies of $M_{\rm gal}\geqslant 10^{10.5}\, \rm M_{\odot}$, we find lower fractions of star-forming galaxies. The difference with the observations may be due to the nature of the void galaxy samples. The SDSS voids being larger, and their samples including many more galaxies, they may have a larger fraction of more massive galaxies than we do, which would result in lower fractions of star-forming galaxies.\\

\subsection{Discussion for the galaxy properties and the assembly of BHs}
Our initial goals were to understand whether the void galaxies have different properties and a different assembly history than galaxies embedded in denser regions and whether these properties were affected by environment. As found in observations, we find that low-mass galaxies are abundant in cosmic voids. Most of these galaxies still lie in the main star-forming sequence of the simulation and we find that on average they are forming stars efficiently with respect to their stellar mass, i.e. have a slightly higher sSFR than the average over all the simulated galaxies of Horizon-AGN. This would indeed favor a ``late'' evolution of the void galaxies, as suggested from observational studies with SDSS \citep{2014MNRAS.445.4045R}. Some of the properties of the BHs also seem to favor a late assembly of the void galaxy population. BHs embedded in voids form later than average BHs in the same mass galaxies in the entire simulation. 
This could indicate that the host galaxies and their BHs assemble later in time in voids than in denser regions. 
In Horizon-AGN, BH sink particles are formed in dense regions with $\rho>\rho_0$, with $\rho_0$ the density threshold. Therefore the seeding of the simulation is closely tied to the assembly of the stellar content of galaxies. 
However, we find that the average stellar mass of the inner void galaxies is smaller than for galaxies in denser regions for the same mass dark matter halos (for $M_{\rm h}\leqslant 10^{11}\, \rm M_{\odot}$). This may indicate that the assembly of galaxies and BHs is more difficult and slower in the inner regions of voids and/or in isolated galaxies of the cosmic web. 

Further investigations are required to confirm that the galaxies present in the small voids studied here did form later on compared to the ones in the dense regions of the simulation. This can be done by looking at their star formation rate history, both in simulations and observations. Further analyses would also need to understand whether this effect also stands for larger cosmic voids (see discussion below).

Most of the BHs located in the inner regions of voids also have a harder time growing, and barely grow above $10^{8}\, \rm M_{\odot}$, for the most efficient. We find that about $20\%$ of the inner BHs could be observed as AGN with Eddington ratio of $\log_{10 }f_{\rm Edd}\geqslant -3$, and hard X-ray luminosities of $L_{\rm 2-10}=10^{41.5-42.5}\, \rm erg/s$. Observing and investigating BH properties in cosmic voids could help us to understand the rise of the BH population at high redshift, to what extent the growth of BHs can be efficient in the absence of galaxy mergers and/or external interaction, and the early co-evolution between BHs and their host properties. In the simulation, we note that the processes by which the void galaxies grow affect the growth of the central BHs in a similar way, as we find that their mass ratios $M_{\rm BH}/M_{\rm gal}$ at $z=0$ is not different from the entire simulation.

The galaxy occupation fraction with BHs, particularly in low-mass galaxies, is a crucial diagnostic of BH formation mechanisms. In the following, we briefly discuss how the occupation fraction could be altered in cosmic voids, and therefore used to improve our understanding of BH formation and evolution.
As most of cosmological simulations \citep[but see][]{Bellovary2011,2017MNRAS.468.3935H}, the seeding of Horizon-AGN does not particularly follow prescriptions from one of the main theoretical BH formation mechanisms (e.g., PopIII remnants, compact stellar clusters, or direct collapse). 
BH sink particles of $\sim 10^{5}\, \rm M_{\odot}$ form in dense gas cells in galaxies where BH formation has not occurred yet. The threshold on gas density tends to disfavor the formation of BHs in low-mass galaxies. These large-scale simulations also lack resolution to resolve accurately galaxies of $M_{\star}\leqslant \rm a\, few\, 10^{8}\, \rm M_{\odot}$. 
Therefore, our findings on the occupation fraction in the crucial low-mass regime in large cosmological simulations should be taken with a grain of salt and will need to be investigated with more adequate methods. We address this in more detail in a forthcoming paper. We just mention here that in Horizon-AGN the occupation fraction of galaxies in voids is slightly different from the occupation fraction average over all the galaxies in the simulation. 
For example, none of the galaxies located at the very center of voids, with $r/R_{\rm void}\leqslant 0.4$, host a supermassive BHs. We also find that the inner voids galaxies with $r/R_{\rm void}\leqslant 0.65$ have a slightly lower occupation fraction than all galaxies. 
Similarly, we find pretty low occupation fractions for isolated galaxies ($V_{\rm gal}\geqslant 10\, \rm (cMpc/h)^{3}$).\\

Can we use cosmic voids to get some hints of BH formation?
Because we find that the BHs in inner void galaxies are still able to growth (at least for some of them) with their galaxies, their pristine properties such as their initial BH masses may have been erased by $z=0$. 

However, we find that these galaxies have a very high fraction of in-situ stellar formation. Therefore these galaxies have a reduced number of galaxy mergers, and so BH-BH mergers (Fig.~\ref{fig:bh_growth}). Without being altered by BH-BH mergers, the BH occupation fraction in the inner cosmic void galaxies could therefore be very close to the true occupation fraction set by the physics of BH formation mechanisms. 
Our results show that more investigations are needed to understand to what extent voids can be used to study BH formation.   

In any case, the fact that these galaxies have lower stellar masses compared to their dark matter halos, makes the inner void galaxies and their BHs very interesting sites to look at for less evolved systems.

\subsection{What about the properties of dark matter halos in cosmic voids? and their assembly with time?}

Here, we have focused our analysis on the properties of galaxies and BHs, but the properties of dark matter halos in cosmic voids could also be different than in denser environments. Several studies have been conducted using cosmological simulations or zoom-in simulations of under-dense and over-dense regions of dark matter halos (and/or galaxies) to study their properties. We note here that the works mentioned below do not always analyze voids but rather low-density regions, and in any case, they do not identify cosmic voids based on a Voronoi tessellation of the tracer distribution (as we do here with galaxies).

Using dark matter simulations, \citet{2017MNRAS.466.3834L} showed that halos in low-density environments are, on average, more elongated, and have lower spins.
The low-density environment of halos does not enclose many other halos that could contribute to building strong enough tidal force fields to spin them up \citep{2017MNRAS.466.3834L}. The correlation between the angular momentum of dark matter halos and the angular momentum of their galaxies is still not clear and debated today \citep[][and references therein]{2019MNRAS.488.4801J}. 
Different halo angular momentum may also involve differences in the angular momentum of their galaxies, their sizes \citep[][and references therein]{2018MNRAS.473.2714S}, and consequently on BH fueling \citep{2019MNRAS.484.4413H}. The interplay between these quantities in cosmic voids needs to be investigated in detail.

Using the Millennium simulation, \citet{2009MNRAS.394.1825F, 2010MNRAS.401.2245F} showed that the halo mass growth of $10^{12}-10^{15}\, \rm M_{\odot}$ halos in void regions was more predominantly obtained through diffuse accretion (instead of mergers). The impact of the environment of these halos was found to be different for different mass bins of dark matter halos.
We emphasize here that the dark matter halos in all these studies are more massive than the halos that we identified in our void sample, due to the smaller volume of the simulation.

Halos in denser regions could have assembled at earlier times than halos in under-dense regions, for same mass halos \citep{2015ApJ...812..104T,2017MNRAS.466.3834L} (for $M_{\rm h}\geqslant 10^{11}\, \rm M_{\odot}$), due to earlier formation times, interaction with neighbors systems at earlier times as well. The presence of more filaments could enhance the feeding of the galaxies embedded in these dense-environment halos more efficiently \citep{2015ApJ...812..104T}. 
Instead, galaxies in low-density environments, and particularly in cosmic voids, could assemble later, in agreement with our findings in this paper. \cite{2015ApJ...812..104T} also shows that while the sSFR of galaxies are higher in under-dense regions at $z=0$, the sSFR were higher at high redshift in dense regions.
A perspective of our work would be to investigate in more detail the mass growth history of dark matter halos and their galaxies in cosmic voids (possible larger voids), to understand the impact of environment in their assembly.

\subsection{Discussion on the properties as a function of void sizes}
In the present paper, we derived the properties of galaxies and BHs as a function of void-centric distances, averaging information from voids with different sizes. Nevertheless, void evolution and features could have a dependence on void size (for details see, e.g., \cite{2004MNRAS.350..517S, Hamaus:2014fma}). Observed void sizes will also change depending on the galaxies that are used as tracers to identify these voids in the first place. In observations, we are limited by the number density of tracers (linked to the minimum stellar mass or magnitude of galaxies that can be detected by the survey). Since the tracer number density in simulations is generally higher than in observations, observed voids can be larger than voids in our study, where we use a galaxy mass limit of a few $10^{8}\, \rm M_{\odot}$. Additionally, voids have a hierarchical structure, hence if a high tracer density is observed by the survey we can find and map in detail the sub-voids composing the larger voids. Our approach to identify voids, based on a Voronoi tessellation and watershed transform, is suitable to measure such a hierarchical structure. It will be of great interest to study the impact on our results of void sizes and void hierarchy with larger simulations, such as IllustrisTNG (300 cMpc length on a side), where the large size provides a robust statistic of voids very different sizes and allows to access the different levels of the hierarchy (voids and sub-voids). The distribution of void sizes in such a simulation and SDSS would be comparable. 
An interesting perspective of our work will be to directly compare the properties of galaxies in observations and such a simulation, possibly with reduced resolution, to mimic the tracer number densities found in observational data.\\

\subsection{Discussion on the void-centric distances vs Voronoi cell volume}
In the paper, we have used two different diagnostics to isolate the void galaxies: their void-centric distances (assuming that the voids have spherical shapes) and the volume of the galaxies' Voronoi cells. 
When considering void-centric distances only, we found that some galaxies are not really located in very low-density regions, but rather in denser regions close to other galaxies. The volume of their Voronoi cells is, indeed, small. Therefore the non-spherical shape\footnote{It has been shown in N-body dark matter simulations that cosmic voids at least larger than $\sim 10 \, \rm Mpc$ radius become more and more spherical with time \citep{1984MNRAS.206P...1I,2004MNRAS.350..517S}, and the shape of voids also becomes increasingly spherical when approaching the interior of voids \citep{1993MNRAS.263..481V}.} of the voids in the simulation affects some of our results when using only the void-centric distances. 
For example, the growth history of some BHs includes rapid growth events. This is likely triggered by external physical processes such as galaxy mergers and dynamical interactions, which are characteristic of non-isolated environments. 
Here we are, of course, limited by low statistics: if more (and more populated) voids were available, it would be possible to perform a more conservative selection on radii, focusing on the very inner part of voids. 
Having a higher number density of tracers would be more reliable and would also allow mapping small voids in detail.
In our work, we find that complementing the void-centric distance approach with the information of the Voronoi cells enhances the robustness of the void selection. It allows to truly select isolated galaxies embedded in the inner regions of cosmic voids.
We plan to investigate the impact of tracer number density with larger simulations that will also provide better statistics for voids with very different features (sizes, tracer number density).
\\

\subsection{Perspectives on observations}
Upcoming observational campaigns will shed new light on the physics of cosmic voids and the evolution of the rare galaxies that they host. The near future surveys of  PFS\footnote{Information on PFS at https://pfs.ipmu.jp/index.html and in \citet{2016SPIE.9908E..1MT}.} ($\rm 15\, deg^{2}$) or WFIRST \citep{2015arXiv150303757S} ($\rm 10\, deg^{2}$ deep field) 
will have high tracer number density, allowing us to study voids of different sizes and relatively high density of tracers even in the very inner regions of voids. 
Broader surveys such as Euclid will have lower resolution but will cover larger field of view, increasing our statistics for the larger voids. 
Interestingly, the new telescope eROSITA \citep{2012arXiv1209.3114M} will soon be able to map the X-ray AGN on the whole sky. It will be possible to compare our results on AGN properties with these new observations, and therefore to better constrain the population of BHs in cosmic voids. This will allow us also to better understand the role of large-scale environment in shaping the properties of BHs.

On the other hand, our computational power increases, and we are now able to simulate larger volume simulations of $300\, \rm  cMpc$ a side, into which we can identify cosmic voids with similar features as the SDSS voids (e.g., their sizes, resolution of the tracers, etc). Among other points, we will also be able to understand the effect of void sizes on the physical properties of galaxies.

In a couple of forthcoming papers, we will pursue our investigations on BH properties in cosmic voids. We will present the results from a dark matter simulation with a larger volume that allows us to study the BH occupation fraction of void galaxies. Finally, we plan to present a more detailed comparison between BH properties from observations in SDSS and simulated void BHs. We will further extend this work using simulations with larger volume and at higher redshift, to make predictions of galaxy and BH properties in cosmic voids in those regimes.

\section*{Acknowledgements}
The authors are grateful to Rien van de Weygaert, Shy Genel, Marta Volonteri and Christophe Pichon for useful discussions.
This work has made use of the Rusty Cluster at the Center for Computational Astrophysics (CCA) of the Flatiron Institute. The Flatiron Institute is supported by the Simons Foundation. 
AP is supported by NASA grant 15-WFIRST15-0008 to the WFIRST Science Investigation Team ``Cosmology with the High Latitude Survey''.
RSS is grateful for the generous support of the Downsbrough family.

\appendix
\section{Growth of BHs in isolated galaxies}
\begin{figure}
\centering
\includegraphics[scale=0.55]{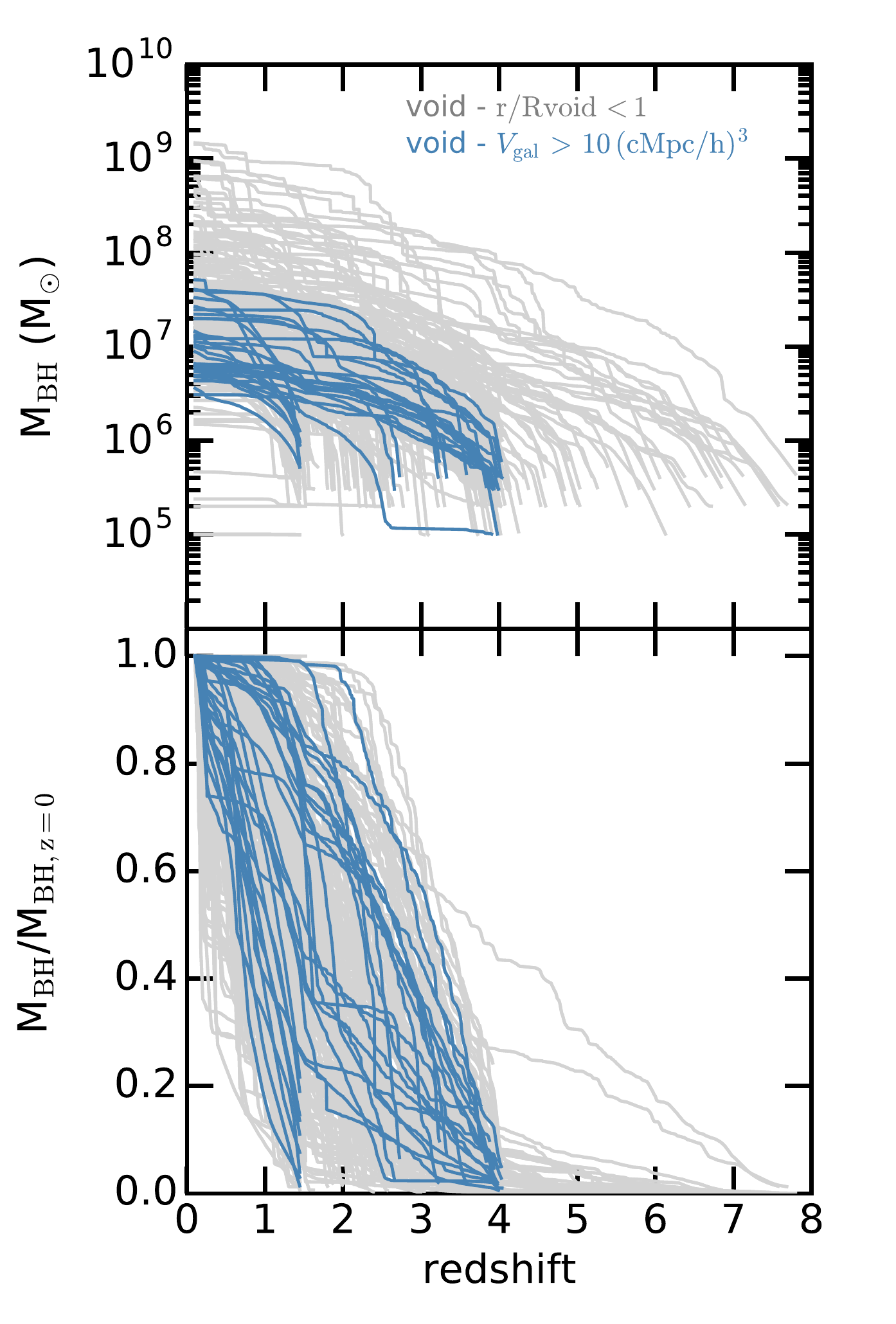}
\caption{BH growth with time, for BHs in void galaxies with $r/R_{\rm void} < 1$ (grey), and isolated galaxies with large Voronoi cell volume of $\rm V_{\rm gal}> 10 (cMpc/h)^{3}$ (blue).}
\label{fig:bh_growth_vol}
\end{figure}

In Fig.~\ref{fig:bh_growth_vol} we show in blue the growth of BHs that are embedded in isolated galaxies at $z=0$. For comparison, grey lines show the growth of BHs in voids ($r/R_{\rm void}<1$). These BHs form more recently and have a hard time growing efficiently to form very massive BHs by the end of the simulation at $z=0$.

\bibliographystyle{mn2e}
\bibliography{biblio_complete}

\label{lastpage}
\end{document}